\documentclass[usenatbib,referee]{mn2e}
\usepackage{amsmath}
\usepackage{graphicx}
\usepackage{epsfig}
\usepackage{txfonts}
\usepackage{color}
\usepackage{natbib}

\title[sdB stars as surviving companions in SNRs] {Subdwarf B stars as possible surviving companions in
Type Ia supernova remnants}
\author[Meng and Li]
{Xiangcun. Meng$^{\rm 1, 2, 3}$ \thanks{E-mail:
xiangcunmeng@ynao.ac.cn}, Jiao. Li$^{\rm 1, 2, 4}$  \\
        $^{1}$ Yunnan Observatories, Chinese Academy of Sciences, 650216 Kunming, PR China\\
$^{2}$ Key Laboratory for the Structure and Evolution of Celestial
Objects, Chinese Academy of Sciences, 650216 Kunming, PR China\\
$^{3}$Center for Astronomical Mega-Science, Chinese Academy of
Sciences, 20A Datun Road, Chaoyang District, Beijing, 100012, P.
R. China\\
$^{4}$University of Chinese Academy of Sciences, Beijing 100049,
China}

\begin{document}
\date{}
\pagerange{\pageref{firstpage}--\pageref{lastpage}} \pubyear{2018}
\maketitle

\label{firstpage}

\begin{abstract}\label{0. abstract}
Although type Ia supernovae (SNe Ia) are so important in many
astrophysical fields, a debate on their progenitor model is still
endless. Searching the surviving companion in a supernova remnant
(SNR) may distinguish different progenitor models, since a
companion still exists in the remnant for the single-degenerate
(SD) model, but does not for the double degenerate (DD) model.
However, some recent surveys do not discover the surviving
companions in the remnant of SN 1006 and Kepler's supernova, which
seems to disfavor the SD model. Such a result could be derived
from an incorrect survey target. Here, based on the
common-envelope wind SD model for SNe Ia, in which the initial
binary system consists of a white dwarf (WD) and a main-sequence
(MS) star, we found that the companion at the moment of supernova
explosion may be a MS, red-giant (RG) or subdwarf B (sdB) star if
a spin-down timescale of less than $10^{\rm 7}$ yr is assumed. We
show the properties of the companions at the moment of supernova
explosion, which are key clues to search the surviving companions
in SNRs. Here, we suggest that the sdB star may be the surviving
companion in some SNRs, even if the progenitor systems are the WD
+ MS systems. The SNe Ia with the sdB companions may contribute to
all SNe Ia as much as 22\%.
\end{abstract}


\section{Introduction}\label{sect:1}
Although Type Ia supernovae (SNe Ia) are so important in
astrophysical fields, e.g. as the best distance indicator to
measure the cosmological parameters (\citealt{RIE98};
\citealt{PER99}; \citealt{MENGXC15}), what their progenitors are
is still an open problem (\citealt{HN00}; \citealt{LEI00}). Now,
it is widely believed that a SN Ia arises from a binary system
with at least one carbon oxygen white dwarf (CO WD)
(\citealt{NUGENT11}; \citealt{WANGB12}; \citealt{MAOZ14}). The
companion of the CO WD may be a normal star, i.e. a main-sequence
or a slightly evolved star (WD+MS), a red giant star (WD+RG) or a
helium  star (WD + He star) (i.e. the single degenerate, SD,
model, \citealt{WI73}; \citealt{NTY84}), or another CO WD
involving the merger of two CO WDs (i.e. the double degenerate,
DD, model, \citealt{IT84}; \citealt{WEB84}). At present, both the
SD and the DD models get supports from observations, but also meet
problems on the observational and theoretical side.

A basic way to distinguish the different models is to search the
surviving companion star in a supernova remnant (SNR), since the
SD model predicts that the companion still exists in the SNR, but
does not for the DD model (but see \citealt{SHENK18}). The
discovery of some potential surviving companions in some SNRs
shows the power of the method (\citealt{RUIZLAPUENTE04};
\citealt{LIC17}). However, many teams also present negative
reports for searching the surviving companions in other SNRs, e.g.
in the two famous Galactic SNRs for SN 1006 and Kepler's
supernova, which seems to favor the DD model
(\citealt{GONZALEZ12}; \citealt{SCHAEFER12};
\citealt{RUIZLAPUENTE17}; \citealt{KEERZENDORF17}). A possible
solution for this embarrassment of the SD model is from the
so-called spin-up/spin-down model, in which the WD is spun up by
accretion, and the rapidly rotating WD experience a long spin-down
phase before supernova explosion. For the long spin-down
timescale, the companion becomes too dim to be detected
(\citealt{JUSTHAM11}; \citealt{DISTEFANO12};
\citealt{BENVENUTO15}). Before supernova explosion, the companion
is usually proposed to become a low-mass helium WD
(\citealt{JUSTHAM11}; \citealt{DISTEFANO12}; \citealt{NOMOTO18}),
but such a suggestion was not confirmed by searching the surviving
WD companion in the remnant of SN 1006 (\citealt{KEERZENDORF17}).
One possible reason is from the huge uncertainty of the spin-down
timescale (\citealt{DISTEFANO11}; \citealt{MENGXC13}). Another
possible solution is from the WD + He star channel, where the
surviving companion is a very luminous helium star, rather than a
dim WD (\citealt{WANGB09}; \citealt{MCCULLY14}), but such a
scenario was not confirmed by numerical simulations and
observations for SN 1006 (\citealt{PANK14};
\citealt{KEERZENDORF12}; \citealt{GONZALEZ12}). Similarly, no MS
or RG stars in the remnant of Kepler's supernova is suitable to be
the surviving companion (\citealt{RUIZLAPUENTE17}). However, could
there be a possibility that the surviving companion is neither a
MS, a RG, a WD, nor a luminous helium star? In this paper, we will
investigate this possibility in details.

In section \ref{sect:2}, we describe our methods and present the
calculation results in section \ref{sect:3}. We show discussions
and our main conclusions in section \ref{sect:4}.

\begin{figure}
\begin{center}
\epsfig{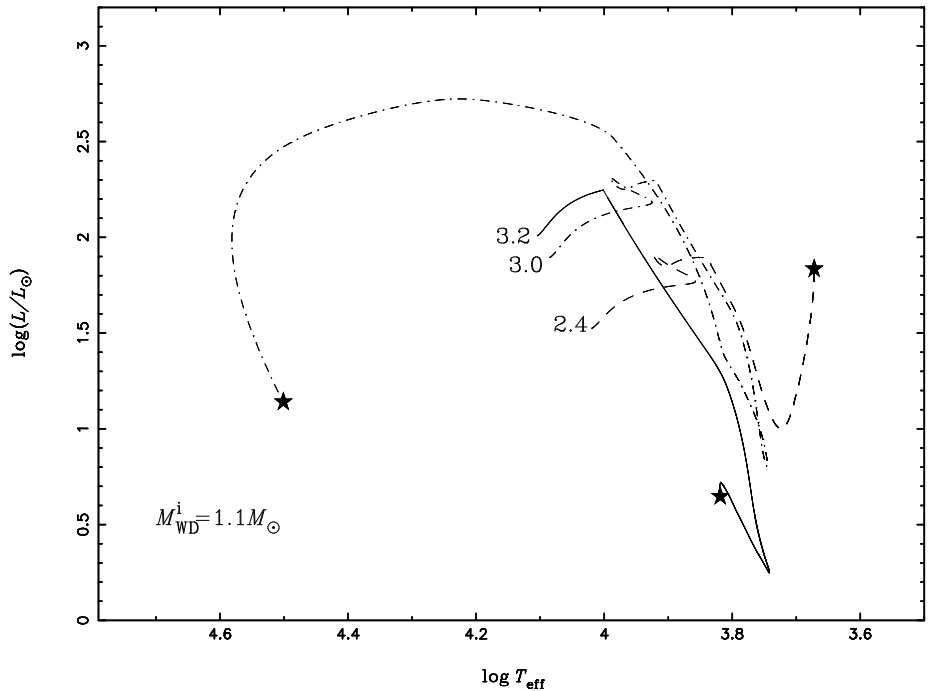}
 \caption{Three examples on the evolution of the companions in HR
diagram, where the initial WD mass $M_{\rm WD}^{\rm
i}=1.1M_{\odot}$, and the initial periods and the initial
companion masses are [$\log (P^{\rm i}/{\rm d}, M_{\rm 2}^{\rm
i}/M_{\odot})$]=(3.2, 0.2), (3.0, 0.5) and (2.4 ,0.5),
respectively. The asterisks denote the position for supernova
explosion.}\label{indihrdsn7}
  \end{center}
\end{figure}

\section[]{METHOD}\label{sect:2}
Recently, \citet{MENGXC17a} developed a new version of the SD
model, i.e. the common-envelope wind (CEW) model which is
completely different from the optically thick wind (OTW) model in
physics (\citealt{HAC96}). In the model, a CO WD accretes
hydrogen-rich material from its companion to increase its mass,
where the companion fills its Roche lobe on the MS or in the
Hertzsprung gap (HG). If the mass-transfer rate between the CO WD
and the companion exceeds a critical accretion rate, a common
envelope (CE) is assumed to form, rather than the onset of the
OTW. Within a very wide parameter range, the binary system in the
CE may avoid merging and lead to a SN Ia explosion finally, where
the SN Ia may explode in the CE phase, in a phase of stable
hydrogen burning or a phase of weakly unstable hydrogen burning.
At present, the CEW model is still under development, and some
parameters used in the present model are relatively conservative,
e.g. the CE density is set to be the average density of the CE by
assuming a spherical CE structure, and the mass-loss rate from the
CE surface is obtained by modifying the Reimer¡¯s wind formula
(\citealt{REIMERS75}). However, the CE model is quite robust and
the CEW model is held within a rather large parameter region (see
the detailed discussions in \citealt{MENGXC17a}).

The WD may obtain a part of the angular momentum of the accreted
material and then rapidly rotate. Since a rapidly rotating WD may
not explode even if its mass exceeds 1.378 $M_{\odot}$
(\citealt{NTY84}; \citealt{YOON04,YOON05}), we continue our
calculations assuming the same WD growth pattern as $M_{\rm WD} <
1.378~M_{\odot}$, since it is not clear how the WD increase its
mass (even not clear whether the WD may continue to increase its
mass) after the time of $M_{\rm WD} = 1.378~M_{\odot}$. To explode
as a SN Ia, the rapidly rotating WD must experience a spin-down
phase (\citealt{JUSTHAM11}; \citealt{DISTEFANO12}), which is very
crucial to show the signature predicted from a single-degenerate
system (\citealt{MENGXC18b}). At present, the spin-down timescale
is quite uncertain in theory (\citealt{DISTEFANO12}). Based on a
fact that circumstellar material (CSM) exists around some SNe Ia,
\citet{MENGXC13} provide a constraint on the spin-down timescale
via a semi-empirical method and they found that the spin-down
timescale should be shorter than a few $10^{\rm 7}$ yr, otherwise
it would be impossible to detect the signature of the CSM. Then,
the spin-down timescale probably between $10^{\rm 5}$ yr and a few
$10^{\rm 7}$ yr (\citealt{DISTEFANO12}; \citealt{MENGXC13}).
However, when is the time of the onset of the spin-down phase is
also uncertain. Here, we assume that the WD explodes as a SN Ia
$10^{\rm 7}$ yr after the WD grows to $M_{\rm
WD}=1.378~M_{\odot}$. Such a timescale may be longer than a real
spin-down timescale by a few $10^{\rm 6}$ yr, i.e. the real
spin-down timescale for the WD is less than $10^{\rm 7}$ yr while
is several times longer than $10^{\rm 6}$ yr, which is consistent
with the empirical constraints (\citealt{MENGXC13,MENGXC18}). We
will discuss the effect of the assumed spin-down timescale on our
results in section \ref{sect:4.1}. We then record the parameters
of the companion at the time of supernova explosion, e.g. the
effective temperature, luminosity, mass, radius, binary period and
so on.

Since a hybrid carbon-oxygen-neon WD could produce some special
SNe Ia, here, we assume that an initial WD leading to SNe Ia may
be as massive as $1.3~M_{\odot}$ (\citealt{CHENMC14};
\citealt{MENGXC14,MENGXC18}). We then calculate a dense model grid
with different initial WD masses, different initial companion
masses and different initial periods. The initial masses of donor
stars, $M_{\rm 2}^{\rm i}$, range from 2.2 to 4.0 $M_{\odot}$; the
initial masses of the WDs, $M_{\rm WD}^{\rm i}$, from 0.8 to 1.30
$M_{\odot}$; the initial orbital periods of binary systems,
$P^{\rm i}$, from $\log(P^{\rm i}/{\rm day})=0.2$ to $\sim15$ day,
at which the companion star fills its Roche lobe at the end of the
HG. The parameter ranges here are smaller than those in
\citet{MENGXC17a,MENGXC18}, since in the other ranges, the
companions appear as MS or RG stars at the moment of supernova
explosion, while we focus on a new kind of surviving companions
that no one has mentioned before.

To investigate the birth rate of SNe Ia with special companions,
we performed two binary-population-synthesis (BPS) simulations as
the method in \citet{MENGXC17a,MENGXC18}. Here, we assume that if
a WD is less massive than 1.3 $M_{\rm \odot}$ and the system is
located in the ($\log P^{\rm i}$, $M_{\rm 2}^{\rm i}$) plane for a
SN Ia at the onset of Roche-lobe overflow (RLOF), a SN Ia occurs.
We followed the evolution of $10^{\rm 7}$ binaries, where the
primordial binary samples are generated in a Monte-Carlo way with
the following input assumptions: (1) a constant star-formation
rate of $5M_{\odot}{\rm /yr}$, or a single starburst of $10^{\rm
11} M_{\odot}$; (2) the initial mass function (IMF) of
\citet{MS79}; (3) a uniform mass-ratio distribution; (4) a uniform
distribution of separations in $\log a$ for binaries, where $a$ is
the orbital separation; (5) circular orbits for all binaries; (6)
a CE ejection efficiency of $\alpha_{\rm CE}=1.0$ or $\alpha_{\rm
CE}=3.0$, where $\alpha_{\rm CE}$ denotes the fraction of the
released orbital energy used to eject the CE\footnote{Note that
the CE here forms on a dynamical time-scale and will also be
ejected on a dynamical time-scale, which is crucial to form some
special close binaries, e.g. the close binaries with hot subdwarf
stars (\citealt{HANZW03}; \citealt{XIONGHR17}), while the CE in
the CEW model is maintained on a thermal time-scale.} (see
\citealt{MENGXC17a} for further details).

To search the surviving companion in a SNR, a color-magnitude
diagram (CMD) is usually used. We may get the color and absolute
magnitude of the companion via interpolation in the empirical
color-magnitude form in \citet{Worthey11}, and then obtain the CMD
of the companion at the moment of supernova explosion. We will
compare our results with the recent surveys in the remnants of SN
1006 and Kepler's supernova (\citealt{KEERZENDORF17};
\citealt{RUIZLAPUENTE17}). For the remnant of SN 1006, we use the
empirical color transformations in \citet{JORDI06} to calculate
the $u-g$ color used in \citet{KEERZENDORF17}. To calculate the
$g$ band apparent magnitude, we adopt a remnant's distance of 2.2
kpc, an extinction of $A_{\rm V}=0.3$ and $A_{g}/A_{\rm V}=1.22$
as in \citet{KEERZENDORF17}. For the remnant of Kepler's
supernova, we adopt a remnant's distance of 5.0 kpc, an extinction
of $A_{\rm V}=2.7$ and $A_{\rm R}/A_{\rm V}=0.748$ as in
\citet{RUIZLAPUENTE17}.

\begin{figure}
\begin{center}
\epsfig{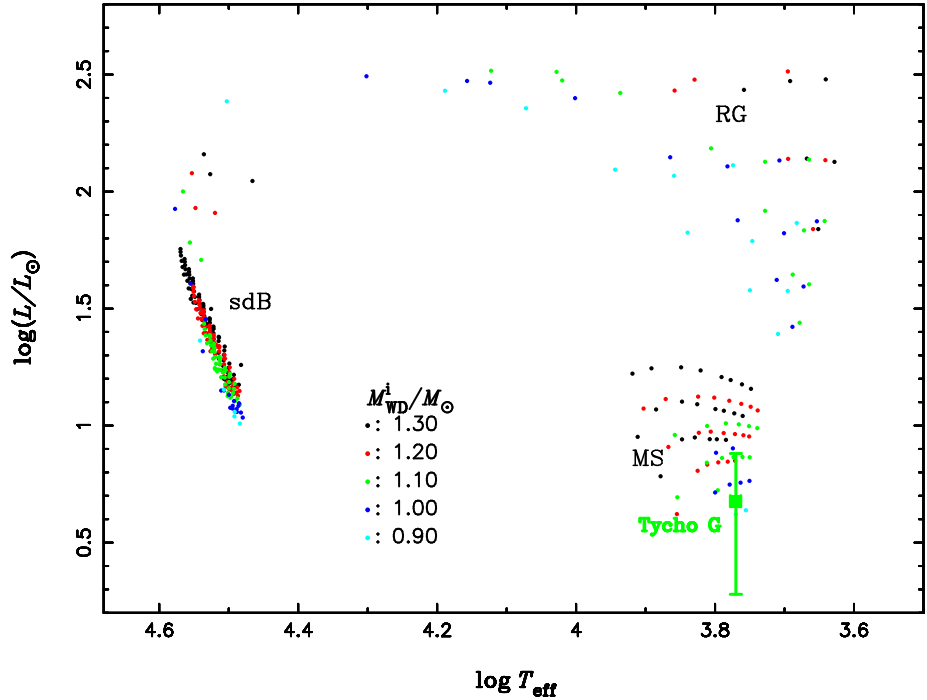}
 \caption{The evolutionary stages of the companions at the moment
of supernova explosion in HR diagram, where different color points
represent different initial WDs. The labels of `MS', `RG' and
`sdB' show the evolutionary states of the companions at the moment
of supernova explosion. Tycho G is a possible candidate of the
surviving companions for Tycho's supernova
(\citealt{RUIZLAPUENTE04}).}\label{hrdsn7}
  \end{center}
\end{figure}

\section{RESULT}\label{sect:3}
\subsection{Different evolutional stages at the moment of supernova explosion}\label{sect:3.1}
In Fig.~\ref{indihrdsn7}, we show the companion evolution for
three different initial binary systems, where the mass transfer
between the WD and the companions begins when the companion is on
the MS or in the HG. Although the initial WD mass is the same for
the three systems, the final evolutionary stages of the companions
at the moment of supernova explosion are quiet different. For the
system that the companion fills its Roche lobe on MS, the
companion is still a MS star at the moment of supernova explosion
for a long nuclear timescale. The other two systems have the same
initial period but different initial companion mass. For the
system with $M_{\rm 2}^{\rm i}=2.4M_{\odot}$, the companion is a
RG star when supernova occurs, while for the system with $M_{\rm
2}^{\rm i}=3.0M_{\odot}$, the companion is a subdwarf B (sdB)
star. Such different destinations are mainly derived from
different mass-transfer rates between the WD and its companions.
For systems with the same initial WD mass and initial orbital
period, the mass-transfer rate for a system with a more massive
companion is higher for a higher mass ratio, i.e. the more massive
companion is more likely to lose its envelope within $10^{\rm 7}$
yr to show the properties of a sdB star.

This result is quite different from previous concept since before
this, a low mass WD was usually expected in a SNR if a spin-down
timescale is considered (see the review by \citealt{NOMOTO18}).
The result here could potentially explain why no surviving
companion was discovered in some Galactic SNRs by the most recent
surveys (\citealt{KEERZENDORF17}; \citealt{RUIZLAPUENTE17}). For
the WD + He star channel, the surviving companions are also helium
star, and the WD + He star channel are supported by some
observations (\citealt{WANGB09}; \citealt{MCCULLY14};
\citealt{GEIER15}), but such helium stars would be much different
from the sdB star predicted here. In the following subsections, we
will show the properties of the companions at the moment of
supernova explosion and show the differences between the sdB stars
here and helium stars from the WD + He star channel.

\begin{figure}
\begin{center}
\epsfig{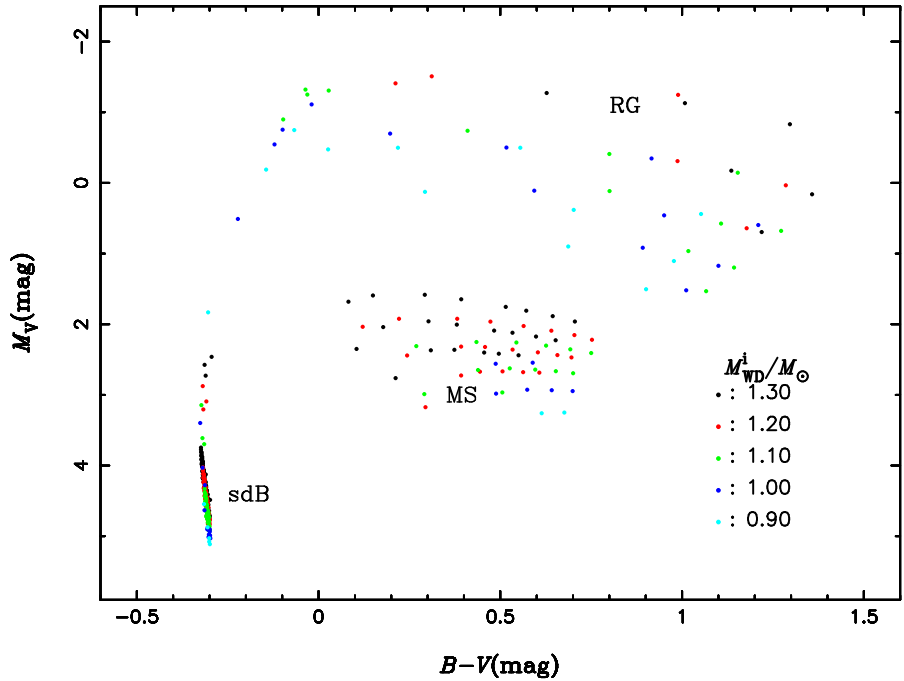}
 \caption{$V$ band absolute magnitude, $M_{\rm V}$, vs. $B-V$ color
for the companions of SNe Ia at the moment of supernova explosion,
where different color points represent different initial WDs. The
labels of `MS', `RG' and `sdB' show the evolutionary states of the
companions at the moment of supernova explosion.}\label{sdbcmd}
  \end{center}
\end{figure}

\subsection{The companions'properties at the moment of supernova explosion}\label{sect:3.2}
\subsubsection{Hertzsprung-Russell and color-magnitude diagrams}\label{sect:3.2.1}
In Fig.~\ref{hrdsn7}, we show the evolutionary stages of the
companions at the moment of supernova explosion in
Hertzsprung-Russell (HR) diagram. In the figure, the companions
are clearly divided into three groups, i.e. the companions may be
MS, RG or sdB stars at the moment of supernova
explosion\footnote{The systems with subgiant stars at the moment
of supernova explosion are rare for their short lifetimes at this
stage, and then we do not take the subgiant stars as a single
group.}. For the MS companion, the effective temperatures are
between 5600 K and 8900 K and the luminosity is between 3
$L_{\odot}$ and 65 $L_{\odot}$, while for the RG stars, the
effective temperatures are between 4000 K and 10000 K and the
luminosity is between 80 $L_{\odot}$ and 320 $L_{\odot}$. However,
the effective temperature of the sdB companions are focus around
$30000\sim40000$ K and the luminosity are between 10 $L_{\odot}$
and 65 $L_{\odot}$. In this paper, we do not calculate the whole
model grids as in \citet{MENGXC17a}, since we just focus on the
SNe Ia with the sdB companions. If all the model grids are
considered, some MS companions may be expected to be dimmer than
0.2 $L_{\odot}$ (see the Fig. 13 in \citealt{MENGXC17a})

The evolutionary state of the companions at the moment of
supernova explosion heavily depends on the evolutionary state of
the companion at the onset of mass transfer and the initial
companion mass. In theory, for a WD + MS system, the companion
fills its Roche lobe on the main sequence or in the HG, and then
mass transfer begins between the WD and the companion star. If the
mass transfer begins when the companion is a MS star, the
companion will be still a MS star at the moment of supernova
explosion for its long nuclear evolution timescale. If the mass
transfer begins when the companion is crossing the HG, the final
evolutionary state of the companion at the moment of supernova
explosion is heavily dependent on the envelope mass upon the
helium core when $M_{\rm WD}=1.378~M_{\odot}$. If the envelope is
too thick to be consumed by mass transfer and nuclear burning
within following $10^{\rm 7}$ yr, it is a RG star at the moment of
supernova. Otherwise, a sdB star is expected. Generally, for a
system with given WD and companion, the longer the initial orbital
period, i.e. the later the mass transfer begins, the more likely
the companion to be a sdB star at the moment of supernova
explosion. On the other hand, for a system with given WD and
orbital period, the more massive the companion, the less massive
the envelope of the companion when $M_{\rm WD}=1.378~M_{\odot}$
(see Figs. 7 and 9 in \citealt{MENGXC17a}) and then more likely
the companion to be a sdB star at the moment of supernova
explosion.

In Fig.~\ref{hrdsn7}, there is a connection between the sdB and
the RG groups, i.e. the sdB stars experience a RG-like stage
before they evolve to the sdB branch. In this paper, the spin-down
timescale is less than $10^{\rm 7}$ yr. It can be imagined that
more companions would be at the sdB phase at the supernova
explosion moment if a longer spin-down timescale is assumed.

In the figure, we also plot a possible candidate of the surviving
companion, i.e. Tycho G, for Tycho's supernova. We can see that
the candidate is consistent with our models with MS companions.
However, for Tycho G, there are still hot arguments on whether or
not it is the surviving companion of Tycho's supernova
(\citealt{RUIZLAPUENTE04}; \citealt{FUHRMANN05};
\citealt{GONZALEZ09}; \citealt{KEERZENDORF09}; \citealt{BEDIN14};
\citealt{XUE15}).

In Fig.~\ref{sdbcmd}, we translate the HR diagram into the CMD
with $V$ band absolute magnitude, $M_{\rm V}$, vs. $B-V$ color for
the companions of SNe Ia at the moment of supernova explosion,
which may provide potential informations for searching the
surviving companion in SNRs. Again, the companions are divided
into MS, RG and sdB groups and there is a connection between the
RG and the sdB groups. However, the CMD presents an important
information, i.e. although most of the sdB stars have a higher
luminosity than the MS ones, their absolute magnitudes in $V$ band
are lower than the MS stars by $0.5$ mag to $3$ mag, which is
derived from their high effective temperatures, i.e. their main
radiations focus on ultraviolet band. So, U or UV band could be an
ideal one to search the sdB companions.

\begin{figure}
\begin{center}
\epsfig{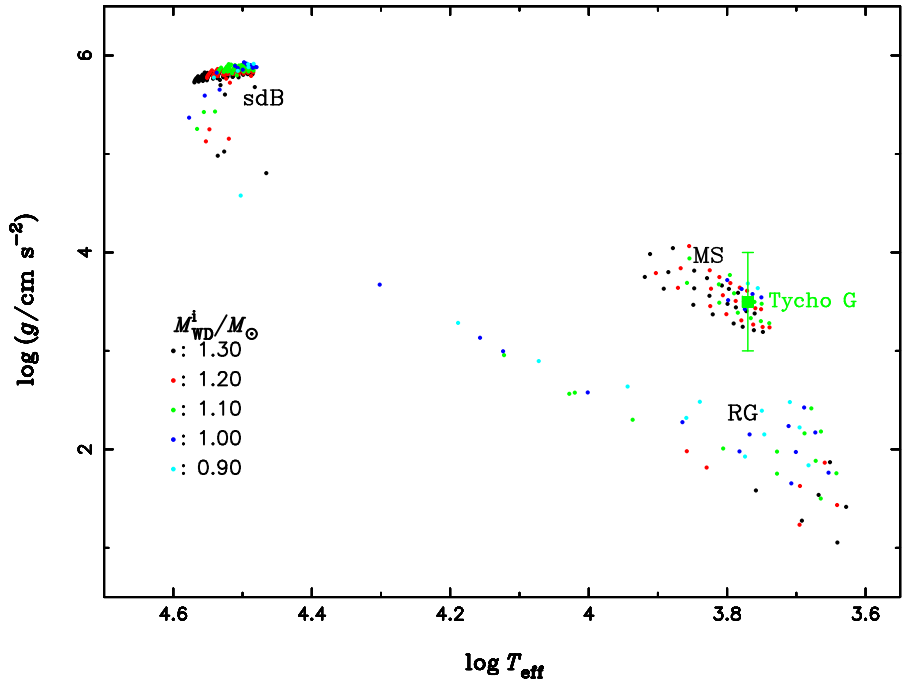}
 \caption{The gravity vs. the effective temperature of the
companions at the moment of supernova explosion, where different
color points represent different initial WDs. The labels of `MS',
`RG' and `sdB' show the evolutionary states of the companions at
the moment of supernova explosion. Tycho G is a possible candidate
of the surviving companion for Tycho's supernova
(\citealt{RUIZLAPUENTE04}).}\label{grasn7}
  \end{center}
\end{figure}

\subsubsection{Gravity}\label{sect:3.2.2}
In Fig.~\ref{grasn7}, we show the gravity vs the effective
temperature of the companions at the moment of supernova
explosion. Again, we see the three groups and the connection
between the sdB and RG groups. The three groups present their
typical surface gravity, i.e. $\log g\sim2$ for RG, $\sim3.5$ for
MS and $\sim6$ for sdB stars.

In the figure, we also plot the possible candidate of the
surviving companion, i.e. Tycho G, for Tycho's supernova. As
discussed above, its surface gravity is also consistent with the
MS group predicted here.

\begin{figure}
\begin{center}
\epsfig{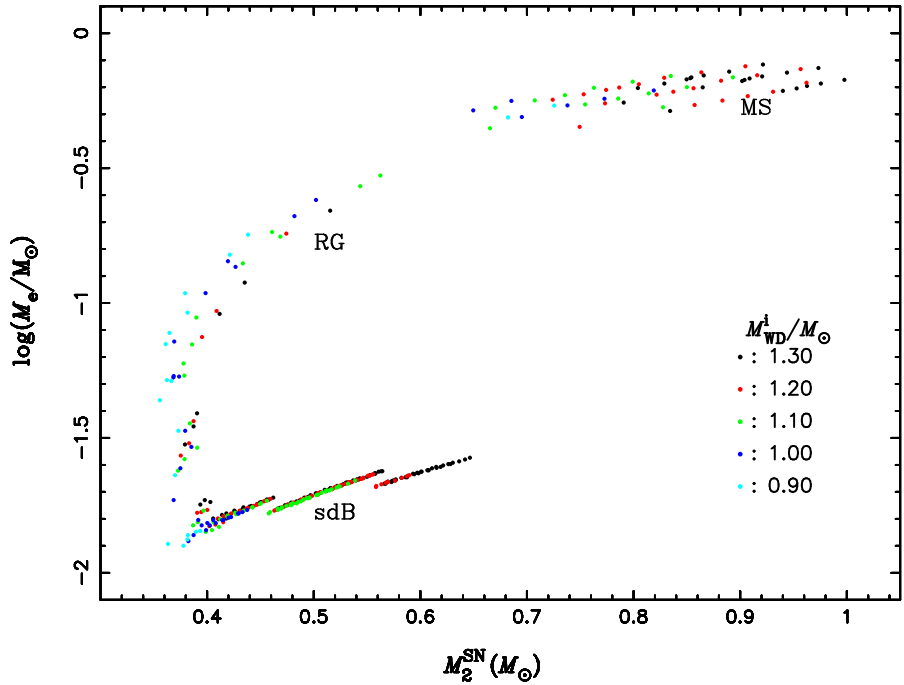}
 \caption{The envelope masses vs. the total masses of the
companions at the moment of supernova explosion, where different
color points represent different initial WDs. The labels of `MS',
`RG' and `sdB' show the evolutionary states of the companions at
the moment of supernova explosion.}\label{mcoresn7}
  \end{center}
\end{figure}

\subsubsection{Mass}\label{sect:3.2.3}
To further present the properties of the companions, we show the
envelope masses and the total masses of the companions at the
moment of supernova explosion (Fig.~\ref{mcoresn7}). Here the
envelope mass, $M_{\rm e}$, is equal to the difference between the
total mass and the core mass, i.e. $M_{\rm e}=M_{\rm 2}^{\rm
SN}-M_{\rm core}^{\rm SN}$, where the definition of the core is
the same to that in \citet{HAN94} and \citet{MENGXC08}. From the
figure, we can see that the envelope mass of sdB companions is
less than 0.03 $M_{\odot}$, but larger than 0.01 $M_{\odot}$, and
their total masses are between 0.4 $M_{\odot}$ and 0.65
$M_{\odot}$, close to the typical mass of sdB stars
(\citealt{HEBER09}). The MS companions have a mass between 0.65
$M_{\odot}$ and 1 $M_{\odot}$. For RG companion, their total
masses are similar to sdB stars, with an envelope of $\sim0.1-0.3$
$M_{\odot}$, while the stars in the connection between the RG and
the sdB stars have an envelope less massive than 0.03 $M_{\odot}$.

After the supernova explosion, the supernova ejecta may impact
into the envelope of the companions and strip off a part of the
envelope (\citealt{MAR00}; \citealt{MENGXC07}; \citealt{PAKMOR08};
\citealt{PANK12}; \citealt{LIUZW12}). So, searching the
stripped-off material in the nebular spectra of SNe Ia is a method
to verify the SD model. However, the deduced amount of the
hydrogen-rich material from the nebular spectra of SNe Ia is
usually smaller than the predicted one from numerically
simulations (\citealt{MAT05}; \citealt{LEO07}; see also
\citealt{MAGUIRE16}). The negative results could be derived from
the uncertainties and limitations on the radiative transfer
calculation (see the discussions in \citealt{MAGUIRE16}). Another
huge uncertainty is that the numerically simulations of the impact
between the supernova eject and its companion do not consider the
effect of the spin-down timescale on the companion structure. The
results here are helpful to explain negative results on searching
the stripped-off hydrogen-rich material in the nebular phase. For
the sdB companion, there is almost no material to be stripped off,
and then their total masses are almost equal to the ones presented
in Fig.~\ref{mcoresn7} after the impact of supernova ejecta. The
MS companions are also compacter than those at the time of $M_{\rm
WD}=1.378M_{\odot}$, and then a less amount of hydrogen material
are expected to be stripped off from the MS companions here
(\citealt{MENGXC07}; \citealt{PAKMOR08}). For the RG stars almost
all the envelope will be stripped off, i.e. $\sim0.1-0.3$
$M_{\odot}$, which is much smaller than previous value obtained by
numerical simulations (\citealt{MAR00}). For the stars in the
connection between RG and sdB stars, there envelopes are also very
thin, and then even if the envelopes were completely stripped off,
it could be difficult to detect them in the nebular spectra of SNe
Ia (see also \citealt{MENGYANG10}).

After the impact of supernova ejecta, the remnant masses of the RG
companions and the ones in the connection between RG and sdB stars
are close to the minimum mass for the helium ignition in the
center of the star (\citealt{HANZW02}). If the helium can be
ignited in the center of the stars, they may also be sdB stars
with a low mass, otherwise luminous helium WDs would be expected
in SNRs (\citealt{JUSTHAM09}; \citealt{MENGYANG10}).

\begin{figure}
\begin{center}
\epsfig{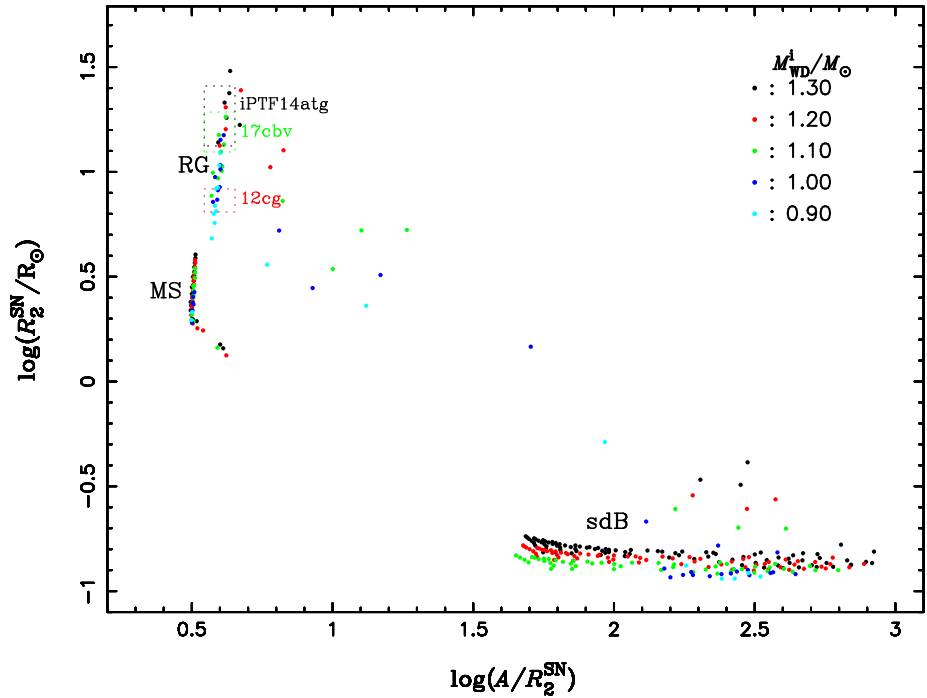}
 \caption{The companion radius and the ratio of the binary
separation to the companion radius at the moment of supernova
explosion, where different color points represent different
initial WDs. The labels of `MS', `RG' and `sdB' show the
evolutionary states of the companions at the moment of supernova
explosion. Three candidates from the SD systems, i.e. iPTF14atg,
SN 2012cg and 2017cbv, are also shown by the dashed rectangles
(\citealt{CAOY15}; \citealt{MARION16};
\citealt{HOSSEINZADEH17})}\label{ar2sn7}
  \end{center}
\end{figure}

\subsubsection{Radius and separation}\label{sect:3.2.4}
The supernova ejecta collides into the companion envelope, and the
kinetic energy of the supernova ejecta is transformed into the
thermal energy of the envelope and finally emits in UV band
(\citealt{KASEN10}). So, detecting the UV excess in the early
light curve of SNe Ia is a powerful method to distinguish
different progenitor models of SNe Ia (\citealt{BROWN12};
\citealt{OLLING15}; \citealt{CAOY15}; \citealt{MARION16};
\citealt{HOSSEINZADEH17}). For a Roche lobe-filling companion, the
luminosity from the collision is proportional to the binary
separation (Eq. 22 in \citealt{KASEN10}). Whether or not the UV
excess may be detected heavily depends on the viewing angle
looking down on the collision region (\citealt{KASEN10};
\citealt{BROWN12}).

In Fig.~\ref{ar2sn7}, we show the companion radius and the ratio
of the binary separation to the companion radius at the moment of
supernova explosion. In the figure, the companions are divided
into two sequences, i.e. the MS and RG companions follow a
sequence with $\log (A/R_{\rm 2}^{\rm SN})\sim0.6$ and the sdB
stars follow the one with $\log (A/R_{\rm 2}^{\rm SN})$ from 1.6
to 3. Such a difference is derived from the fact that in our
model, most of the MS and RG companions are still filling their
Roche lobes at the moment of supernova explosion, while the mass
transfer between the WD and its companion stops for sdB companions
before supernova explosion. For MS companions, there is a tail, in
which the systems are detached. Similarly, the systems in the
connection between the RG and the sdB groups are also detached.
From the figure, we can conclude that the energies received by the
sdB stars from the kinetic energy of the supernova ejecta is lower
than those MS and RG companions by a factor of $10^{\rm 2}$ to
$10^{\rm 5}$, i.e, the UV excess from the collision between
supernova ejecta and the companion can not be detected if the
companion is a sdB star at the moment of supernova explosion,
whatever the viewing angle is. This result is helpful to explain
why so few SNe Ia show signs of UV excess among the hundreds of
SNe Ia (see also the discussion in \citealt{MENGXC16}).

In Fig.~\ref{ar2sn7}, we also present three candidates of SNe Ia
from the SD systems, i.e. iPTF14atg, SN 2012cg\footnote{There is
still argument on whether or not SN 2012cg is from the SD system
(\citealt{SCHAPPEE18}).} and 2017cbv (\citealt{CAOY15};
\citealt{MARION16}; \citealt{GRAUR16}; \citealt{HOSSEINZADEH17}).
All the three SNe Ia are consistent with our RG companions. This
is a natural result since the larger the separation, the high the
luminosity from the collision for a Roche lobe-filling companion
(\citealt{KASEN10}). For a similar value of $A/R_{\rm 2}^{\rm SN}$
between MS and RG companion, the larger the companion radius, the
high the luminosity from the collision. As shown in
Fig.~\ref{ar2sn7}, the MS companions have a radius between $1.2
R_{\odot}$ and $4.5 R_{\odot}$, while the RG companions have a
radius between $6.0 R_{\odot}$ and $32 R_{\odot}$. So, it is more
likely to detect the UV excess from the system with a RG companion
than the one with a MS companion. In \citet{MARION16}, the SN
2012cg is suggested to be from a SD system with a MS star of
$6M_{\odot}$. Actually, based on our detailed binary evolution, no
any companion is as massive as $6M_{\odot}$ at the moment of
supernova explosion (see also \citealt{MENGXC17a}). From
Fig.~\ref{ar2sn7}, the progenitor of SN 2012cg is more likely to
be a low-mass RG companions if the system is from a SD system (see
also \citealt{BOEHNER17}). The sdB companions usually have a
radius from $0.1 R_{\odot}$ to $0.2 R_{\odot}$, a typical value
for sdB stars (\citealt{HEBER09}).

\begin{figure}
\begin{center}
\epsfig{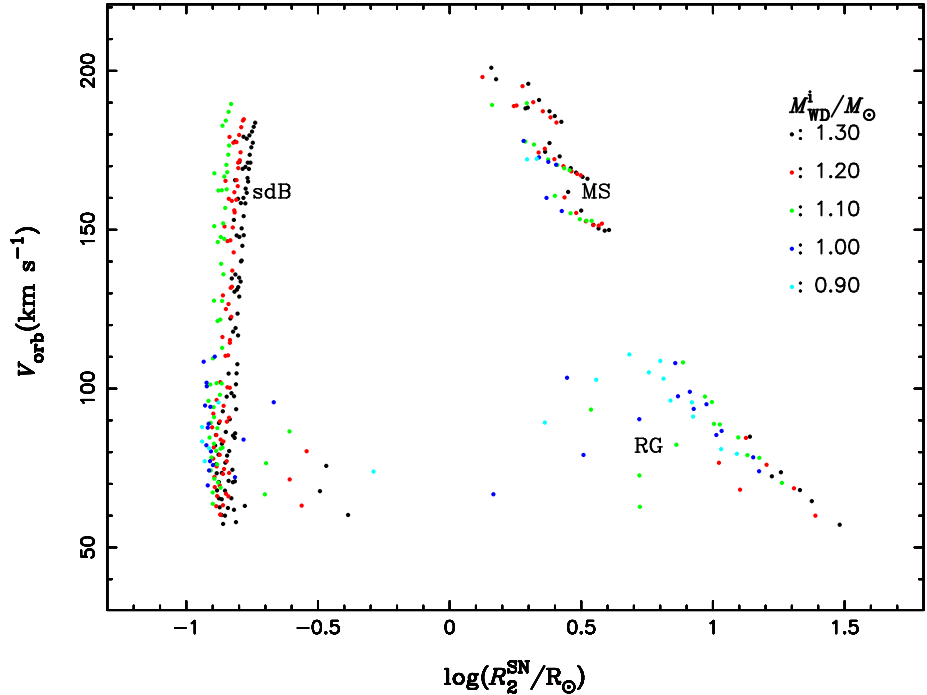}
 \caption{The companion orbital velocity relative to binary
centroid vs. the companion radius at the moment of supernova
explosion, where different color points represent different
initial WDs. The labels of `MS', `RG' and `sdB' show the
evolutionary states of the companions at the moment of supernova
explosion.}\label{vorr2sn7}
  \end{center}
\end{figure}

\begin{figure}
\begin{center}
\epsfig{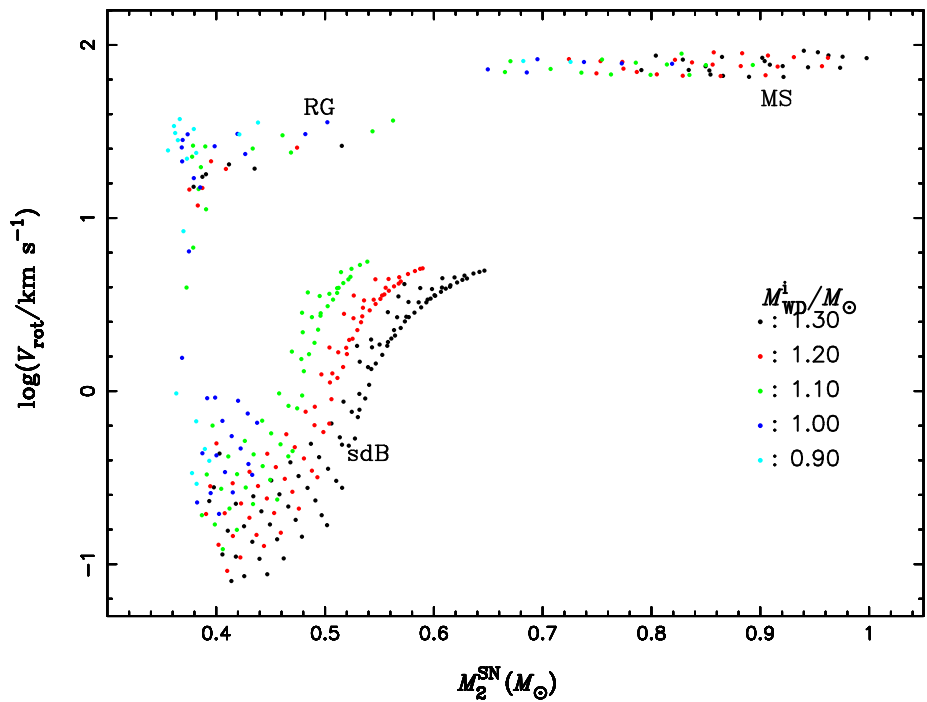}
 \caption{The equatorial rotational velocity of the companion vs.
the companion mass at the moment of supernova explosion, where
different color points represent different initial WDs. The labels
of `MS', `RG' and `sdB' show the evolutionary states of the
companions at the moment of supernova explosion.}\label{vrotm2sn7}
  \end{center}
\end{figure}

\subsubsection{Orbital and rotational velocities}\label{sect:3.2.5}
To judge the origin of a star in a SNR, its kinetics
characteristics may provide very important informations.
Generally, except for being stripped-off a part of its envelope,
the companion may receive a velocity kick from the supernova
ejecta, but the kick velocity is usually much smaller than the
orbital velocity (\citealt{MAR00}; \citealt{MENGXC07};
\citealt{PAKMOR08}; \citealt{PANK12}; \citealt{LIUZW12}). Then,
the orbital velocity of the companion at the moment of supernova
explosion may represent its final space velocity to a great
extant, especially for the sdB companions here, which almost do
not receive any kick velocity for their large value of $A/R_{\rm
2}^{\rm SN}$. If the spatial velocity of a star in a SNR is very
different from the others in the SNR, the star is very possible to
be the surviving companion in the remnant
(\citealt{RUIZLAPUENTE04}).  In Fig.~\ref{vorr2sn7}, we present
the companion orbital velocity relative to binary centroid vs. the
companion radius at the moment of supernova explosion. From the
figure, we can see that different companions may have very
different orbital velocity. For MS companions, the orbital
velocity is from 150 ${\rm km~s^{\rm -1}}$ to 200 ${\rm km~s^{\rm
-1}}$, while the RG companions have an orbital velocity of 50
${\rm km~s^{\rm -1}}$ to 110 ${\rm km~s^{\rm -1}}$. For the sdB
companions, the orbital velocity covers a large range, from 50
${\rm km~s^{\rm -1}}$ to 190 ${\rm km~s^{\rm -1}}$. Such a large
range is mainly derived from the large initial orbital period
range for the systems producing sdB companions, i.e. $\log (P^{\rm
i}/{\rm d})$ is from $\sim0.4$ to 1.2. In other words, although
the mass transfer between a binary system must begin in HG for
producing a sdB companion, it may occur at the stage very close to
the MS end or at the end of the HG. The upper limit of the orbital
velocity here is lower than that in \citet{MENGXC17a} by 60 ${\rm
km~s^{\rm -1}}$, which originates from the fact that after $M_{\rm
WD}=1.378M_{\odot}$, the binary orbital period increases with mass
transfer for a reversed mass ratio (see Fig. 2 in
\citealt{MENGXC17a}). For the same reason, the lower limit of the
orbital velocity is also lower than that in \citet{MENGXC17a}.

In Fig.~\ref{vrotm2sn7}, we show the equatorial rotational
velocity of the companion at the moment of supernova explosion,
where the equatorial rotation velocity is calculated by assuming
that the companion star co-rotates with the orbit\footnote{For a
single RG star, conservation of angular momentum requires a faster
rotating cores than its envelope, which could mean a fast rotating
WD from the RG star (\citealt{BECK12}). However, the observed sdB
stars are generally slowly rotating, which could be from their
much larger radii by a consideration of the conservation of
angular momentum (\citealt{HEBER09,HEBER16}).}. It is clearly
shown in the figure that the MS companions have the fastest
rotational velocity between 60 ${\rm km~s^{\rm -1}}$ and 100 ${\rm
km~s^{\rm -1}}$, and the RG companions have a rotational velocity
of 10 to 40 ${\rm km~s^{\rm -1}}$. For the same reason mentioned
in the above paragraph, the rotational velocity of the MS
companions here is much lower than that in \citet{MENGXC17a} by
about 100 ${\rm km~s^{\rm -1}}$. Especially, since the RG
companions do not rotate very fast, the rotational velocity could
not be a good clue to exclude a RG star as the surviving companion
in a SNR. For their small radii and long orbital periods, the
rotational velocity of the sdB companions is lower than 6 ${\rm
km~s^{\rm -1}}$, as low as 0.1 ${\rm km~s^{\rm -1}}$. So, the
rotational velocity can not be taken as a clue to exclude a sdB
star as the surviving companion in a SNR. The rotational velocity
even might not be a good clue to exclude the MS star as the
surviving companion since the collision between supernova ejecta
and the companion may significantly reduce the rotational velocity
of the companion (\citealt{MENGYANG11}; \citealt{PANK12};
\citealt{LIUZW13}).

\begin{figure}
\begin{center}
\epsfig{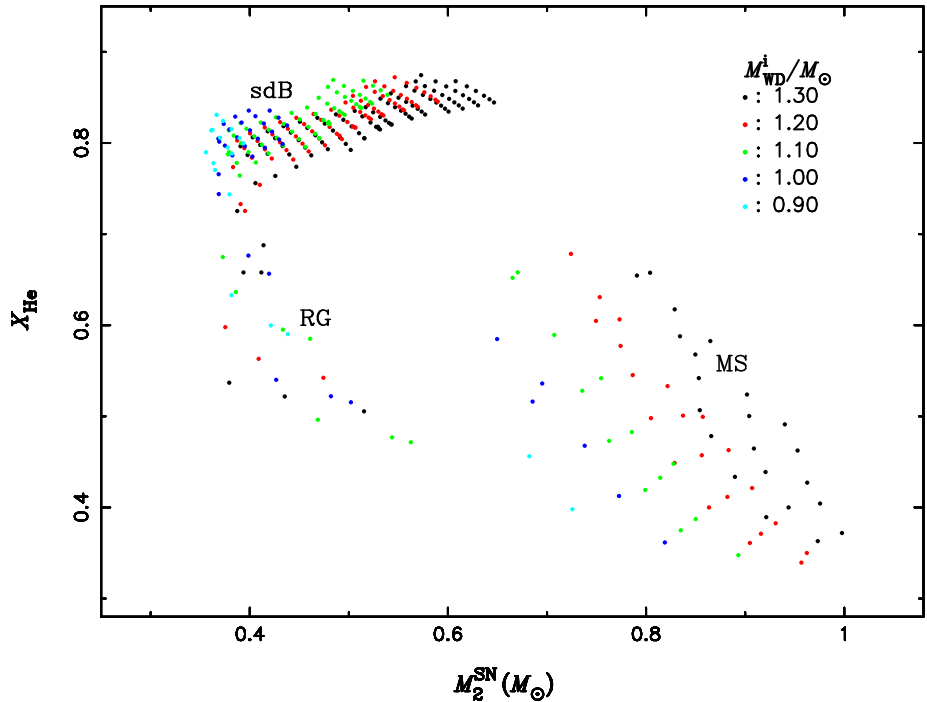}
 \caption{The surface helium abundance of the companion vs. the
companion mass at the moment of supernova explosion, where
different color points represent different initial WDs. The labels
of `MS', `RG' and `sdB' show the evolutionary states of the
companions at the moment of supernova explosion.} \label{xhem2sn7}
  \end{center}
\end{figure}

\subsubsection{Surface helium abundance}\label{sect:3.2.6}
For the SD model, the surface hydrogen-rich material of the
companion is transferred onto the WD, and then the surface
material of the companion at the moment of supernova explosion may
become relative hydrogen-poor. In Fig.~\ref{xhem2sn7}, we present
the surface helium abundance of the companion vs. the companion
mass at the moment of supernova explosion, which may be a key clue
to diagnose whether or not a star in a SNR is the surviving
companion, especially for MS or RG companion. We can see from the
figure that all the companions are significantly helium-enriched,
no matter what the companion type is, i.e. the MS, RG and sdB
companions have a surface helium abundance between 0.35 and 0.7,
between 0.45 and 0.83, and between 0.75 and 0.88, respectively.
Except for the mass transfer between the WD and its companion, a
RG may be still expected to have a higher helium abundance than a
normal MS star via the first dredge up, but its surface helium
abundance can not be as high as $\sim0.8$. Therefore, if a single
RG star in a SNR has a significantly higher helium abundance than
other RG stars, it could be surviving companion in the SNR.
Similarly, a high surface helium abundance of a single MS star in
a SNR increases the possibility of the MS star as the surviving
companion in the SNR. For example, the properties of an evolved
MS-like star in the center region of N103B are similar to the
prediction in \citet{PODSIADLOWSKI03}. If it is the surviving
companion, it could have a high surface helium abundance. In
addition, the sdB stars here have an intermediate surface helium
abundance, while those in close binaries from the canonical CE
ejection channel have either a very high or a very low surface
helium abundance (\citealt{XIONGHR17}).

\begin{figure}
\begin{center}
\epsfig{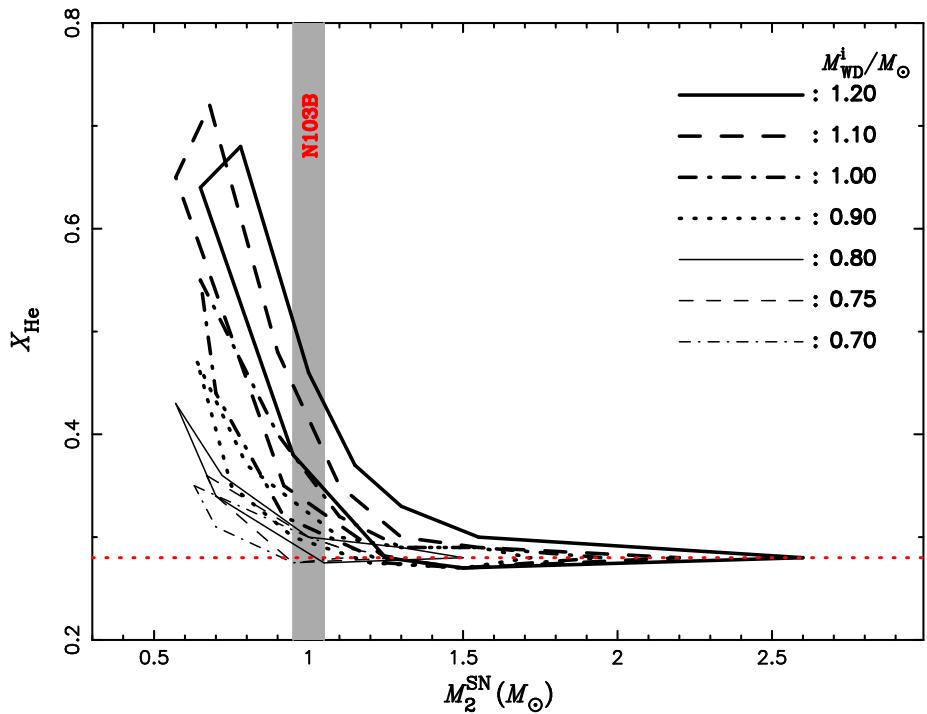}
 \caption{The contour of the surface helium abundance of the
companion vs. the companion mass when $M_{\rm WD}=1.378M_{\odot}$
for different initial WD masses, where the data are from
\citet{MENGXC17a}. The dotted line shows the initial value of the
helium abundance of the companions. The shaded region shows the
possible value of the progenitor candidate in the SNR N103B, where
the companion mass is assumed to be $1M_{\odot}$ based on the
simulation in \citet{PODSIADLOWSKI03}.} \label{xhem2}
  \end{center}
\end{figure}

As a comparison, we also show the surface helium abundance of the
companions when $M_{\rm WD}=1.378M_{\odot}$. From the figure, we
can see that many systems may still produce the companions with
the high surface abundance, i.e. the companions with a mass of
larger than $1.5M_{\odot}$ still have an initial value of the
helium abundance, while for the companions with a mass of less
than $1M_{\odot}$, the helium abundance is larger than its initial
value. In addition, a massive initial WD is more likely to lead to
the companions with the high surface helium abundance, and then
some companions with a mass of $1M_{\odot}$ to $1.5M_{\odot}$ when
$M_{\rm WD}=1.378M_{\odot}$ also have a high surface helium
abundance for the systems with the initial massive WDs. The stars
with the high surface helium abundance are from those that
companions fill their Roche lobe in HG, and will become the sdB or
RG companions if there is a spin-down timescale. So, whatever the
spin-down timescale is, there are some companions with the high
surface helium abundance at the moment of supernova explosion.
Therefore, if a star in a SNR has a significantly higher surface
helium abundance than other stars in the SNR, the star is very
likely to be the surviving companions in the SNR. In
Fig.~\ref{xhem2}, we also show the possible region of the helium
abundance for the candidate of the surviving companion in N103B,
assuming a mass of $1M_{\odot}$ based on the simulation in
\citet{PODSIADLOWSKI03}. It is clearly shown in the figure that
the candidate would have a surface helium abundance larger than
its initial value if it is the surviving companion in N103B, even
though no spin-down timescale is assumed. Then, if the future
spectra observation may verify that the surface helium abundance
of the candidate is significantly higher than the other stars in
N103B, the star is very likely to be the surviving companion.

\subsubsection{The differences from the WD + He star channel}\label{sect:3.2.7}
Here, our model predicts that if there is a delay time of a few
$10^{\rm 6}$ yr for the rapidly rotating WD to explode, the
companion may be a sdB star even if the initial binary system is a
WD + MS system, i.e. the surviving companion in a SNR may be a sdB
star. Whatever, a sdB companion is not a new result since the
surviving companion may also be a helium star for the WD + He star
channel (\citealt{WANGB09a}; \citealt{MCCULLY14};
\citealt{GEIER15}). However, the sdB companions predicted here are
quite different from those from the WD + He star channel. In the
followings, we summarize the main differences between the sdB
companions here and those from the WD + He star channel.

1) There is not any hydrogen on the surviving companion from the
WD + He star channel, while a thin hydrogen envelope on the sdB
stars here is expected.

2) For the existence of the hydrogen envelope, the effective
temperature of the companions here is much lower than that from
the WD + He star channel, e.g. the effective temperature of the
companions from the WD + He star channel is higher than
$5\times10^{\rm 4}$ K, even higher $10^{\rm 5}$ K
(\citealt{WANGB09}), but it is less than $4\times10^{\rm 4}$ K
here. In other words, the companions from the WD + He star channel
will show the properties of extreme helium-rich subdwarf O (sdO)
stars. For example, both US 708 and J2050, the suggested surviving
companions of SNe Ia from the WD + He star channel, are such
extreme helium-rich sdO stars (\citealt{GEIER15};
\citealt{ZIEGERER17}).

3) Before a companion star evolves to the sdB branch, the star
experiences a RG-like phase (see the connection between sdB and RG
groups in Fig.~\ref{hrdsn7}), i.e. the separation of the binary
system here is much larger than that from the WD + He star
channel. Especially, the system becomes a detached binary system,
rather than with a Roche-lobe filling companion. Then, the orbital
velocity and the rotational velocity of the companion here are
much lower than those from the WD + He star channel, i.e. the
orbital velocity of the sdB companion here is smaller than 190
${\rm km~s^{\rm -1}}$, but it is larger than 280 ${\rm km~s^{\rm
-1}}$, and its spacial velocity is higher than 350 ${\rm km~s^{\rm
-1}}$, even as high as 1200 ${\rm km~s^{\rm -1}}$, for the helium
star from the WD + He star channel (\citealt{WANGB09};
\citealt{GEIER15}).

4) Similarly, the rotational velocity of the sdB companions here
is lower than 6 ${\rm km~s^{\rm -1}}$, but it is larger than 140
${\rm km~s^{\rm -1}}$ for the helium star from the the WD + He
star channel (\citealt{WANGB09}).

5) The sdB companions here have a typical mass of sdB stars, i.e.
from $0.4 M_{\odot}$ to $0.65 M_{\odot}$, while the helium star
from the WD + He star channel is more massive than $0.6
M_{\odot}$, even as massive as $2.0 M_{\odot}$
(\citealt{WANGB09}). For the different masses and different
evolutionary stages, the surface gravity of the sdB companions
here is usually higher than that of the helium star from WD + He
star, i.e. the sdB companions here have a surface gravity of $\log
g\sim6$, while the surface gravity of the helium stars from WD +
He star is usually lower than 6, even as low as 4.5
(\citealt{WANGB09}).

6) After the supernova explosion, the supernova ejecta may inject
into the companion envelope, and then the helium companion may
become very luminous, e.g. the luminosity of the helium star
companion may be as high as $10^{\rm 4}L_{\odot}$
(\citealt{PANK14}). However, the solid angle of the sdB companion
to the explosion center is very small here, i.e. about $1/2000$ to
$1/10^{\rm 6}$ of the supernova ejecta is injected into the sdB
companion, and then the supernova ejecta almost can not
significantly change the brightness of the sdB companions. In
other words, the brightness of a sdB companion in a SNR does not
significantly deviate from its brightness at the moment of
supernova explosion. Even the impact of the supernova ejecta were
not considered, the surviving helium star from the WD + He star
channel is still much brighter than the sdB companion here, e.g.
the helium star may have a luminosity of $10^{\rm 2}L_{\odot}$ to
$10^{\rm 5}L_{\odot}$, but the luminosity of the sdB companions
here is only from $10L_{\odot}$ to $65L_{\odot}$
(\citealt{WANGB09}; \citealt{WANGB14b}). Except for a lower mass,
such a luminosity difference is mainly derived from the fact that
the sdB companions here are central-helium-burning, while the
central helium for most of the helium star from the WD + He star
channel is exhausted, i.e. they are helium RG stars
(\citealt{WANGB09}; \citealt{WANGB14b}).

\begin{figure}
\begin{center}
\epsfig{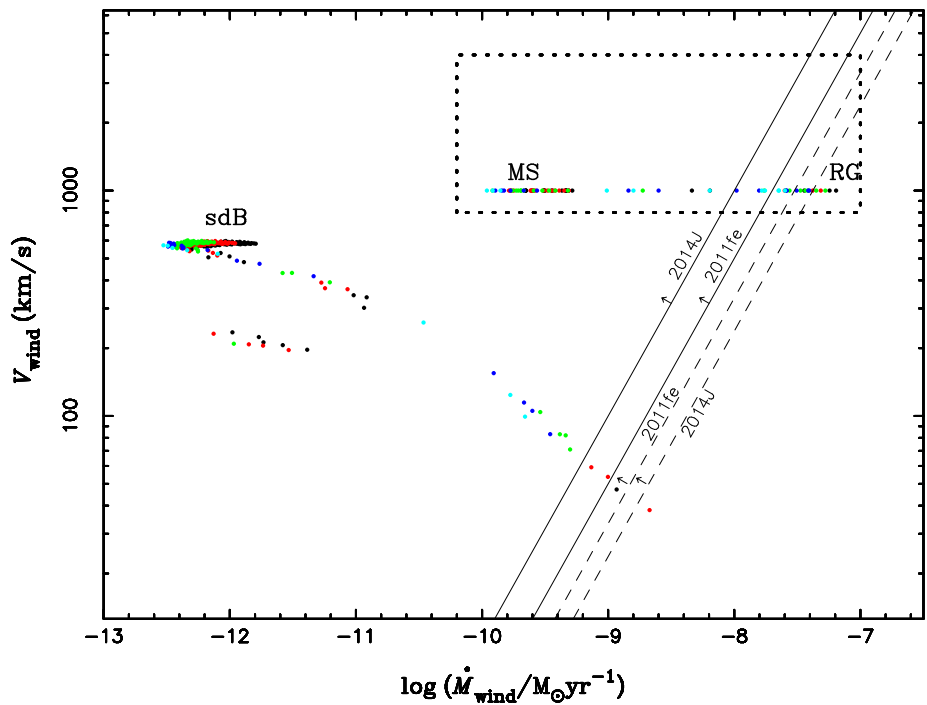}
 \caption{Wind velocity vs. mass-loss rate from the SD system at
the moment of supernova explosion, where different color points
represent different initial WDs. The labels of `MS', `RG' and
`sdB' show the evolutionary states of the companions at the moment
of supernova explosion. The dotted rectangle shows the possible
uncertainty region for wind velocity from WD surface. The left
region of the solid and the dashed lines represents the permitted
region for SD systems constrained by X-ray and radio observations
on SN 2011fe and SN 2014J, respectively
(\citealt{MARGUTTI12,MARGUTTI14}; \citealt{CHOMIUK12};
\citealt{PEREZ14}).} \label{windsn7}
  \end{center}
\end{figure}

\subsection{The mass loss rate from the binary system at the moment of supernova explosion}\label{sect:3.3}
For a SD system, some material may lose from the binary system by
wind to form circumsteller material (CSM). To search the CSM is
then a key method to distinguish from different progenitor models,
e.g. SN 2006X is suggested from a SD system for the discovery of
the variable circumstellar Na I absorption lines in its spectra
(\citealt{PATAT07}). If the CSM is not faraway from the explosion
center, after the supernova explosion, the explosion ejecta runs
into the CSM and interacts with it. If the amount of the CSM is
large enough, a narrow hydrogen emission line may be observed in
the spectra of the SN Ia (\citealt{HAMUY03};
\citealt{SILVERMAN13}). Such SNe Ia are called as SNe Ia-CSM and
are suggested to be from the hybrid CONe WD + MS systems
(\citealt{MENGXC18}). The emission at radio and X-ray bands is
also expected from the interaction between supernova ejecta and
the CSM (\citealt{CHEVALIER90}). Therefore, radio and X-ray
observations may be wonderful tool to diagnose the progenitor
origin of SNe Ia by shedding light on the properties of the CSM
since CSM are generally not formed from the DD model (but see
\citealt{SHENKJ13} for a different view; \citealt{BOFFI95};
\citealt{ECK95}). For the two well observed SNe Ia, 2011fe and
2014J, all of the spectral, radio and X-ray observations indicate
that their environments are very clear (\citealt{NUGENT11};
\citealt{MARGUTTI12,MARGUTTI14}; \citealt{CHOMIUK12};
\citealt{PEREZ14}). In addition, no luminous object is discovered
at the position of these SNe Ia in their archive images before
supernova explosion (\citealt{LIWD11}; \citealt{KELLY14}). These
observations seem to support that SN 2011fe and 2014J originate
from the DD systems.

In Fig.~\ref{windsn7}, we show the wind velocity vs. mass-loss
phase space of the binary system at the moment of supernova
explosion. In the figure, we also plot the constraints from the
radio and X-ray observations for SN 2011fe and 2014J
(\citealt{MARGUTTI12,MARGUTTI14}; \citealt{CHOMIUK12};
\citealt{PEREZ14}). In our model, for a delay timescale of a few
$10^{\rm 6}$ yr, the WDs stop increasing their masses although
there is a mass transfer between the WDs and the RG companions.
Mass transfer does not stop for most of the systems with the MS
companions, either. However, the mass transfer stops for the sdB
companions and the stars in the connection between the RG and the
sdB groups. For the systems with the mass transfer, the material
transferred onto the WDs is lost from the surface of the WDs, and
then we assume a wind velocity of $1000~{\rm km}^{\rm -1}$ with an
uncertainty range from $800~{\rm km}^{\rm -1}$ to $4000~{\rm
km}^{\rm -1}$ (the dotted rectangle). If the mass transfer stops,
the material is lost from the companion surface by a Reimers¡¯s
wind (\citealt{REIMERS75}), where we assume a wind velocity of a
half of the escape velocity of the companions.

It is clearly shown in Fig.~\ref{windsn7} that for the systems
without mass transfer, their mass loss rate is very low, even
lower then $10^{\rm -12} M_{\odot}{\rm /yr}$, i.e. the environment
around such SNe Ia is very clear. For those with mass transfer,
the environment around SNe Ia with MS companions is also relative
clear, but not as clear as those with the sdB companions. The CSM
around the SNe Ia with the RG companions may be dense enough to be
detected by present radio and X-ray observation. Of course, the
spectral observation may also detect the CSM around the SNe Ia
with RG companions. This result may explain why the SNe Ia with
the CSM tend to be from the WD + RG systems (\citealt{PATAT07};
\citealt{DILDAY12}). From Fig.~\ref{windsn7}, we can see that even
for the most well observed SN 2011fe and 2014J, our results still
leave rather large parameter range for the SD model. Especially,
if the companion stars of SN 2011fe and SN 2014J are sdB or
low-mass MS stars, their clear circumstance (\citealt{NUGENT11};
\citealt{MARGUTTI12,MARGUTTI14}; \citealt{CHOMIUK12};
\citealt{PEREZ14}) and negative results to search the stripped-off
hydrogen-rich material in their nebular phase
(\citealt{SCHAPPEE11}; \citealt{SCHAPPEE13};
\citealt{LUNDQVIST15}) and negative results for searching the
progenitor systems in their archive images (\citealt{LIWD11};
\citealt{KELLY14}) may be simultaneously explained. Based on the
study in \citet{MENGXC18b}, the initial WD masses of SN 2011fe and
2014J tend to be relatively small, while the progenitors producing
the sdB companions usually have a massive initial WD. So, the
companions of SN 2011fe and 2014J are more possible to be the
low-mass MS stars if they are from the SD systems (see also
\citealt{BROWN12b}).

In Fig. \ref{windsn7}, there is tail for the sdB stars, which are
those in the connection between the sdB and the RG groups in
Fig.~\ref{hrdsn7}. For this systems, the mass transfer stops and
the envelope of the companions is very thin (see
Fig.~\ref{mcoresn7}). There is another small group under the sdB
companions, which consists of the MS companions without filling
their Roche lobe (see also the MS tail in Fig.~\ref{ar2sn7})

\begin{figure}
\begin{center}
\epsfig{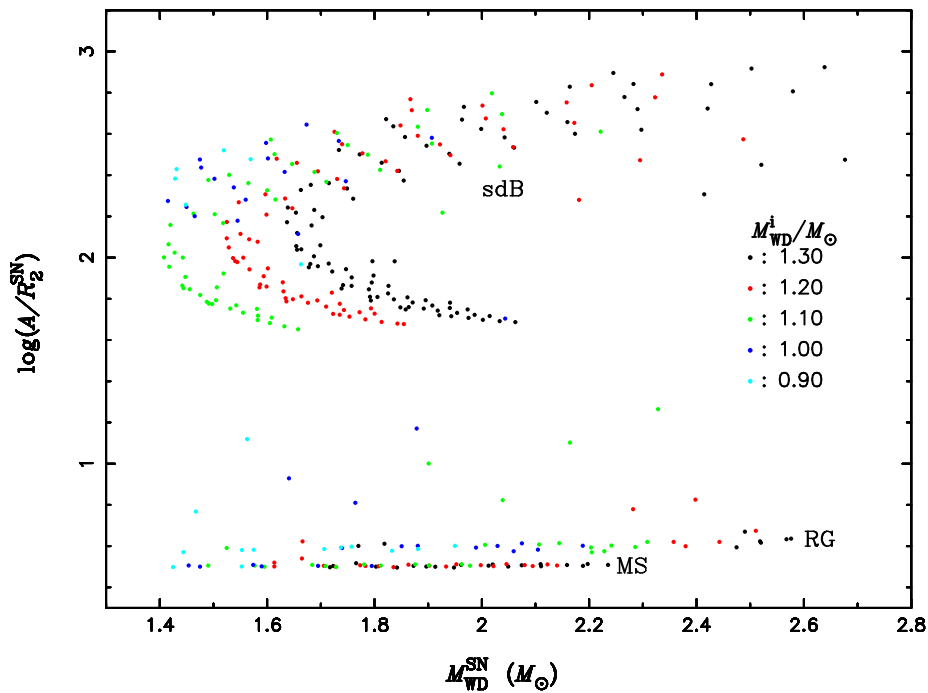}
 \caption{The ratio of binary separation to the secondary radius
vs. the WD mass at the moment of supernova explosion, where
different color points represent different initial WDs. The labels
of `MS', `RG' and `sdB' show the evolutionary states of the
companions at the moment of supernova explosion.} \label{awdsn7}
  \end{center}
\end{figure}

\subsection{The WD mass at the moment of supernova explosion}\label{sect:3.4}
The WD mass at the moment of supernova explosion could play a key
role on the final brightness of a SN Ia (\citealt{HACHISU12};
\citealt{WANGB14}). The WD may exceeds 1.378 $M_{\odot}$ if it
rotates rapidly (\citealt{NTY84}; \citealt{YOON04,YOON05}), and
the mass of the WD depends on its rotational pattern. For pure
rigid rotation, the WD may be as massive as $\sim1.5~M_{\odot}$,
and the the WD with a mass of larger than $1.5~M_{\odot}$ is
differential rotating (\citealt{SAIO04}; \citealt{YOON04,YOON05};
\citealt{HACHISU12a}). For the extreme differential rotation case,
the WD may even be stable as massive as $4~M_{\odot}$
(\citealt{OSTRIKER68}). Here, instead of solving the detailed
stellar structure of a WD, we assume the same WD growth pattern as
$M_{\rm WD} < 1.378~M_{\odot}$ as did in \citet{WANGB14} (see also
\citealt{CHENWC09}; \citealt{HACHISU12a}), and then, we may obtain
the final WD mass at the moment of supernova explosion. In
Fig.~\ref{awdsn7}, we show the ratio of binary separation to the
secondary radius vs. the WD mass. From the figure, we can see that
the WD mass covers a large range, i.e. the WD mass may be as
massive as 2.7 $M_{\odot}$. In addition, the maximum final mass
that a WD may reach is determined by its initial mass, i.e. the
more massive the initial WD, the more massive the maximum final
mass. However, it must be emphasized that the WD mass shown in
Fig.~\ref{awdsn7} should be taken as an upper limit, which is
derived from our assumption. For example, after $M_{\rm
WD}>1.378~M_{\odot}$, the thermonuclear nuclear luminosity may
exceed the Eddington luminosity of the WD for the assumption here,
and then a super-Eddington wind is expected, which may reduce the
timescale that the system stay in the CE phase, and then the final
WD mass (see \citealt{MENGXC17a}).

The rapidly rotating WDs with super-Chandrasekhar mass are
suggested to be the progenitor candidate of super-luminous SNe Ia
(\citealt{CHENWC09}; \citealt{HACHISU12a}; \citealt{WANGB14}).
However, there are still many uncertainties on this subject.
Firstly, it is generally accepted that the maximum brightness of a
SN Ia is determined by the amount of $^{\rm 56}$Ni produced during
the explosion, but the final production of $^{\rm 56}$Ni for the
rapidly rotating WDs is heavily dependent on whether or not a
detonation may be ignited in the WD
(\citealt{PFANNES10a,PFANNES10b}; \citealt{FINK18}). Even a
detonation is ignited, the explosion could not fit the properties
of super-luminous SNe Ia, i.e. it's difficult to reproduce the low
ejecta velocity of super-luminous SNe Ia
(\citealt{PFANNES10a,PFANNES10b}; \citealt{FINK18}). Secondly, as
seen from the Fig.~\ref{awdsn7}, a high initial WD mass is
necessary to produce a significant super-Chandrasekhar mass WD
(see also \citealt{WANGB14}). However, a high initial mass WD may
be an hybrid CONe WD, rather than a CO WD
(\citealt{DENISSENKOV13}; \citealt{CHENMC14}). The explosion of
the hybrid CONe WDs are suggested to produce sub-luminous
2002cx-like SNe Ia, rather than super-luminous SNe Ia
(\citealt{MENGXC14,MENGXC18}; \citealt{WANGB14b}). In addition,
for some super-luminous SNe Ia, a super-Chandrasekhar mass WD
could not be necessary (\citealt{SCALZO12}) and even for those
showing the properties of super-Chandrasekhar mass progenitor, it
is more possible to be from a merger of two massive CO WDs
(\citealt{TAUBENBERGER13}). As discussed above, it is still
unclear which kind of SNe Ia associate with the rapidly rotating
super-Chandrasekhar-mass WD predicted here.

\begin{figure}
\begin{center}
\epsfig{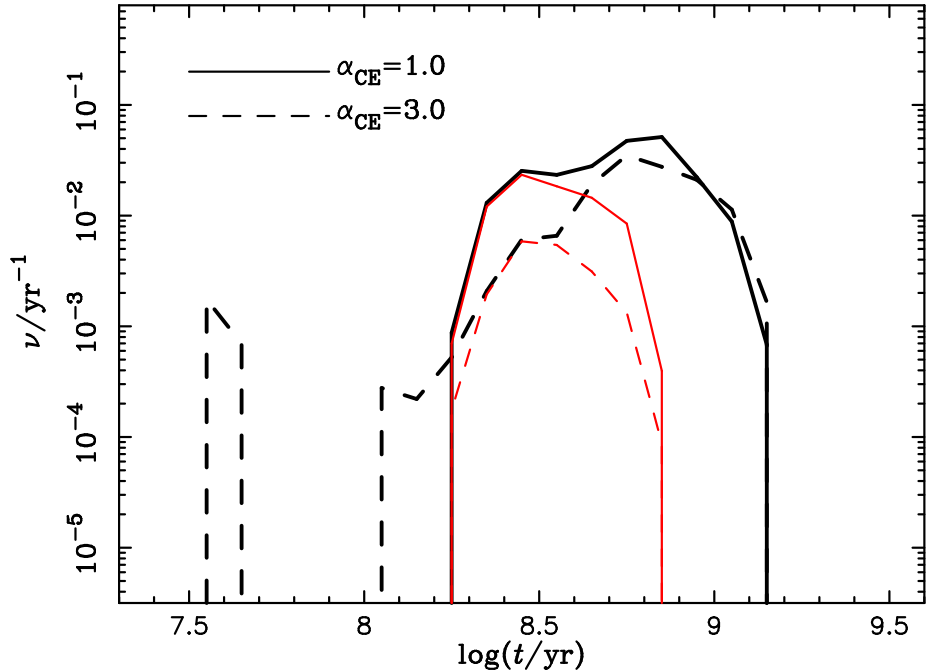}
 \caption{The evolution of the birth rates of SNe Ia for a single
starburst of $10^{\rm 11}M_{\odot}$. The solid and dashed curves
show the cases with $\alpha_{\rm CE}=1.0$  and $\alpha_{\rm
CE}=3.0$, respectively. The black lines show the total birth rate
from the WD + MS channel (\citealt{MENGXC17a,MENGXC18}), while the
red lines show the cases with sdB companions.} \label{singlesn7}
  \end{center}
\end{figure}

\begin{figure}
\begin{center}
\epsfig{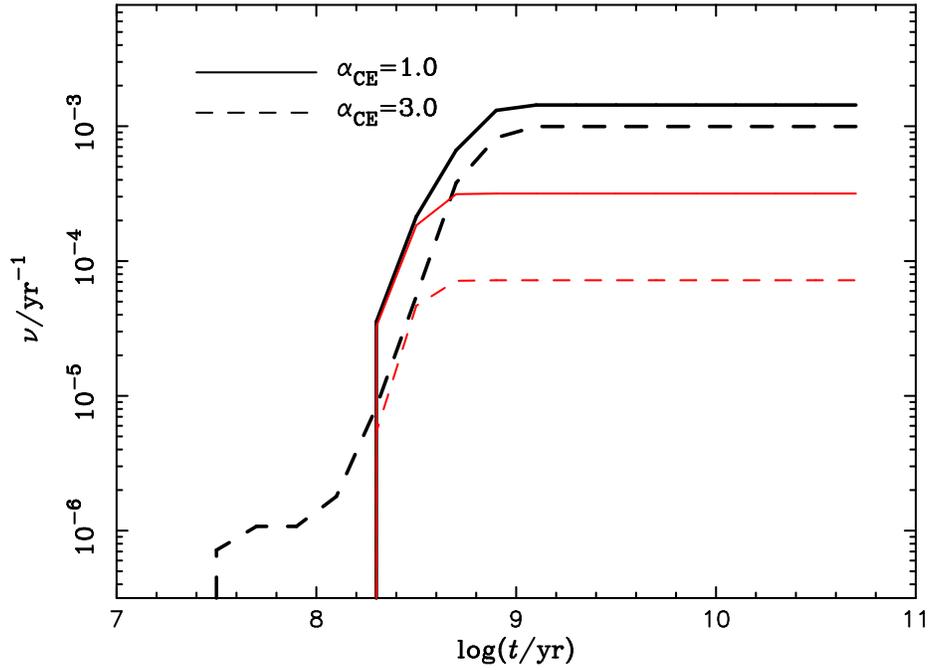}
 \caption{The evolution of the birth rates of SNe Ia for a constant
SFR of $5M_{\odot}{\rm /yr}$. The solid and dashed curves show the
cases with $\alpha_{\rm CE}=1.0$  and $\alpha_{\rm CE}=3.0$,
respectively. The black lines show the total birth rate from the
WD + MS channel (\citealt{MENGXC17a,MENGXC18}), while the red
lines show the cases with sdB companions.} \label{sfrsn7}
  \end{center}
\end{figure}

\subsection{The birth rate of SNe Ia with the sdB companions}\label{sect:3.5}
In this paper, we suggest for the first time here that the
companion of a SN Ia may be a sdB star, although its progenitor
system is a WD + MS one. This is a quiet new result since a low
mass WD is usually expected in a SNR if a spin-down timescale is
considered (\citealt{NOMOTO18}). The question is how common the
sdB companions are for SNe Ia. Fig.~\ref{singlesn7} show the
evolution of the birth rate of SNe Ia for a single starburst,
where the black lines show the total birth rate from the WD + MS
channel, while the red lines show the cases with sdB companions.
Most SNe with the sdB companions occur between 160 and 800 Myr
after the starburst and the birth rate peaks at $\sim300$ Myr. As
expected, the birth rate of SNe Ia with the sdB companions is
lower, but not much lower than the total birth rate of SNe Ia. For
the black dashed line, there is an alone spike at $\sim40$ Myr,
which is from the He star channel, as defined in \citet{MENG09},
in which primordial primary fulfills its Roche lobe in the HG or
on the RG branch (see also \citealt{MENGXC14}).

Fig.~\ref{sfrsn7} shows the Galactic birth rates of SNe Ia from
the WD + MS systems, where the red lines are for SNe Ia with the
sdB companions. The birth rate of SNe Ia with the sdB companions
is between $7.5\times10^{\rm -5}{\rm yr^{\rm -1}}$ and
$3.2\times10^{\rm -4}{\rm yr^{\rm -1}}$, i.e. the SNe Ia with the
sdB companions roughly contribute to 7.3\% to 22\% of all SNe Ia
from the WD + MS channel, depending on the $\alpha_{\rm CE}$
value. As discussed in section~\ref{sect:3.2.1}, if the spin-down
timescale is longer than the value assumed here, the contribution
of SNe Ia with the sdB companions will become higher.

\begin{figure}
\begin{center}
\epsfig{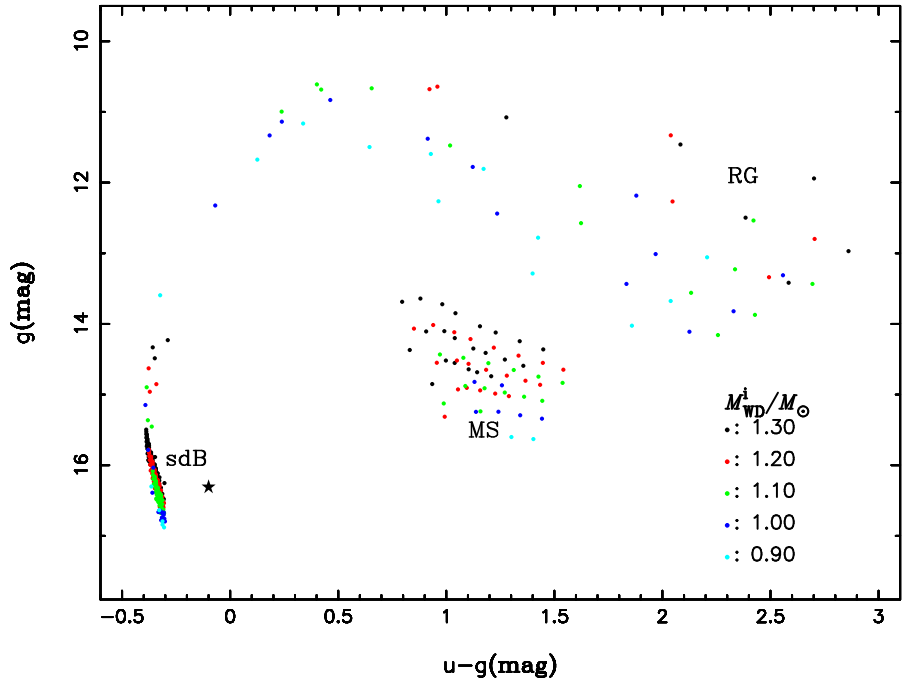}
 \caption{Similar to Fig.~\ref{sdbcmd}, but for $g$ band apparent
magnitude vs. $u-g$ color for the companions of SNe Ia at the
moment of supernova explosion for SN 1006, where different color
points represent different initial WDs. The labels of `MS', `RG'
and `sdB' show the evolutionary states of the companions at the
moment of supernova explosion. The star shows a possible candidate
of the surviving companion for SN 1006 (\citealt{KEERZENDORF17}).}
\label{sdbcmdug}
  \end{center}
\end{figure}

\begin{figure}
\begin{center}
\epsfig{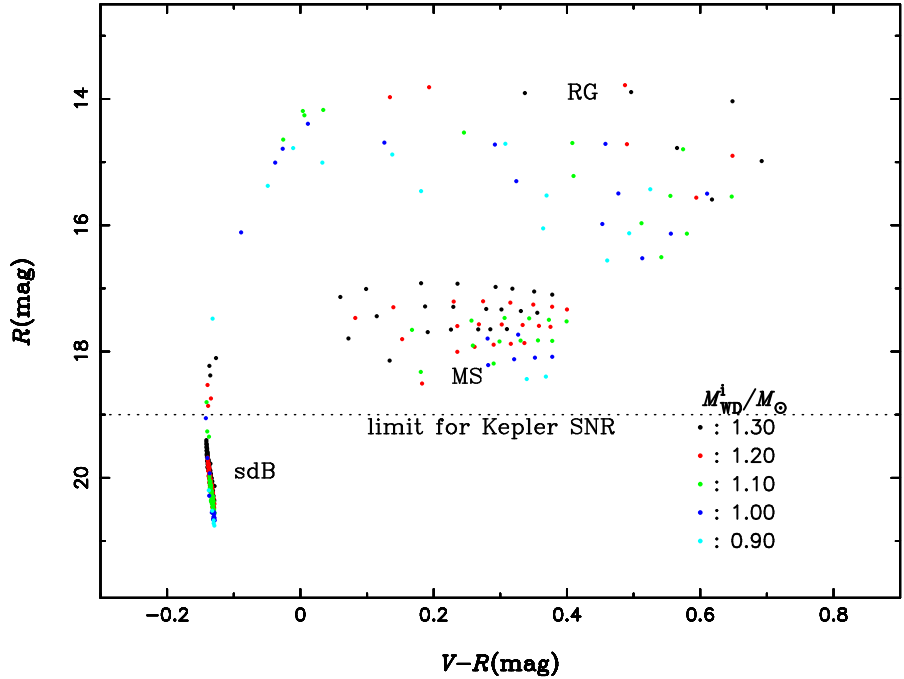}
 \caption{Similar to Fig.~\ref{sdbcmd}, but for $R$ band apparent
magnitude vs. $V-R$ color for the companions of SNe Ia at the
moment of supernova explosion for Kepler's supernova, where
different color points represent different initial WDs. The labels
of `MS', `RG' and `sdB' show the evolutionary states of the
companions at the moment of supernova explosion. The dotted line
shows the limiting apparent magnitude \textbf{of the $R$ band} in
the survey of \citet{RUIZLAPUENTE17}.} \label{sdbcmdvr}
  \end{center}
\end{figure}

\subsection{Comparison with the survey in the remnants of SN 1006 and Kepler's suprnova}\label{sect:3.6}
In this section, we directly compare our result with the survey
results in the remnants of SN 1006 and Kepler's supernova. The CMD
of the surviving companions predicted by our model for SN 1006 is
shown in Fig.~\ref{sdbcmdug}, which is similar to
Fig.~\ref{sdbcmd} but for $g$ band apparent magnitude vs. $u-g$
color. The observation in \citet{KEERZENDORF17} is deep enough to
detect the surviving companion of SN 1006 if it is a MS or RG star
predicted here, but neither a MS nor a RG star in the remnant was
suggested to be the surviving companion (\citealt{GONZALEZ12};
\citealt{KEERZENDORF12}). A sdB star could be an interesting
alternative. Interestingly, a candidate in the sample of
\citet{KEERZENDORF17} is quite similar to our predicted sdB stars,
with a slightly redder color (the star in Fig.~\ref{sdbcmdug}). If
the candidate is the surviving companion of SN 1006, the redder
color would be partly derived from reddening by dust and the
collision between supernova ejecta and the companion
(\citealt{PODSIADLOWSKI03}; \citealt{PANK14}, see discussions in
section~\ref{sect:4.2}).

The main X-ray features of the remnant of Kepler's supernova may
be well explained by a symbiotic binary with a white dwarf and an
asymptotic giant branch (AGB) star, where the interaction between
the stellar wind from the AGB star and the interstellar material
plays a key role to form the present shape of the remnant
(\citealt{CHIOTELLIS12}). Then, the surviving companion of
Kepler's supernova would be a luminous AGB or post-AGB star, but
no such candidate was discovered (\citealt{KEERZENDORF14}).
Similarly, the X-ray image of the N103B also needs an AGB
progenitor, but a MS-like star in the N103B center is suggested to
be \textbf{the} surviving companion (\citealt{LIC17}). Actually,
based on the CEW model, a WD + MS system may also explain the
X-ray features of the remnants, in which the CEW replaces the
stellar wind of the AGB star. Especially, the CEW model may
explain the X-ray image of the N103B and MS-like surviving
companion, simultaneously.

The survey in the remnant of Kepler's supernova by
\citet{RUIZLAPUENTE17} has a limiting apparent magnitude $m_{\rm
R}=+19$ mag, and the photometry and proper motions are complete
down to $m_{\rm F814W}\simeq22.5$, where F814W band is similar to
a wide I band. In Fig.~\ref{sdbcmdvr}, we show the $R$ band
apparent magnitude vs. $V-R$ color for the surviving companions
assuming a distance to Kepler's supernova and a visual extinction
in the direction to the remnant of Kepler's supernova. Again, the
figure clearly shows that the survey of \citet{RUIZLAPUENTE17}
would detect the surviving companion of Kepler's supernova if it
is a MS or a RG star. No MS or RG star is suitable to be the
surviving companion of Kepler's supernova
(\citealt{RUIZLAPUENTE17}). However, the survey focuses on the red
bands and cannot exclude the sdB stars as the surviving companion
of Kepler's supernova, e.g. almost all the sdB stars in
Fig.~\ref{sdbcmdvr} are dimmer than the detection limit of $R$
band in \citet{RUIZLAPUENTE17}. If a sdB star is the surviving
companion of Kepler's supernova, our model may simultaneously
explain the X-ray features of the remnant and the negative reports
on searching the surviving companion in the remnant. If someone
wants to check whether or not a sdB star is the surviving
companion of Kepler's supernova, the survey must be as deep as
$m_{\rm R}\sim+21$ mag. Obviously, both $R$ and F814W bands have a
too long wavelength and is not a good choice to search the sdB
star in the remnant of Kepler's supernova, while $U$ or $UV$ band
would be an excellent alternative. Whatever, a low-mass MS
surviving companion is still possible for the Kepler's supernova
since the low-mass MS companion may be as dim as 0.2 $L_{\odot}$
(\citealt{MENGXC17a}), which is much lower than the detection
limit in \citet{RUIZLAPUENTE17} by one order of magnitude.

\section{DISCUSSIONS AND CONCLUSIONS}\label{sect:4}

\begin{figure}
\begin{center}
\epsfig{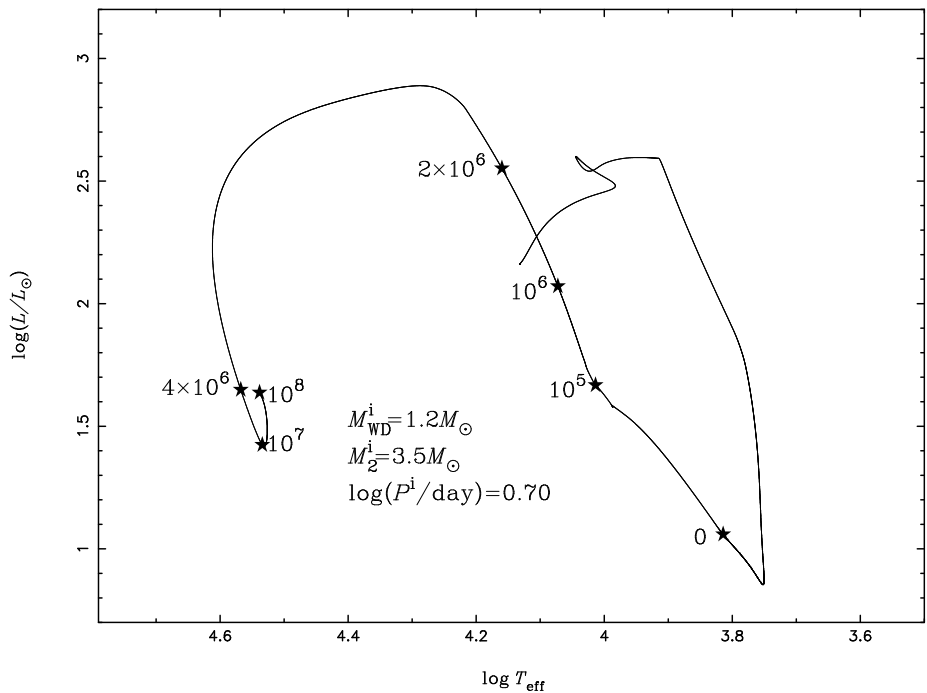}
 \caption{The evolution of the companion for the initial binary
system of [$M_{\rm WD}^{\rm i}/M_{\odot}$, $M_{\rm 2}^{\rm
i}/M_{\odot}$, $\log (P^{\rm i}/{\rm d})$]=(1.2, 3.5, 0.7) with
different spin-down timescales.} \label{hrdspsn}
  \end{center}
\end{figure}

\begin{figure}
\begin{center}
\epsfig{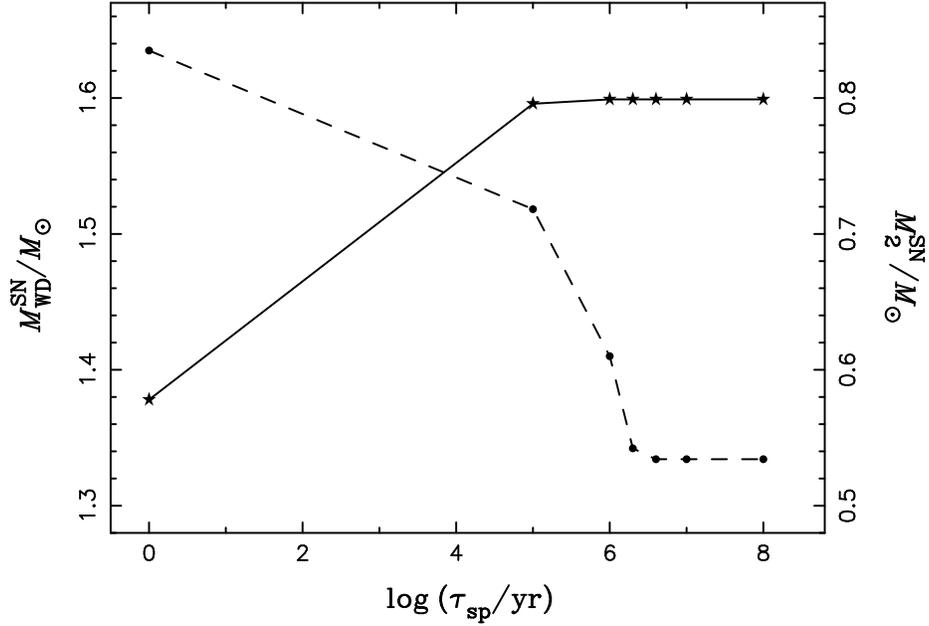}
 \caption{The final WD and companion mass for the initial binary
system of [$M_{\rm WD}^{\rm i}/M_{\odot}$, $M_{\rm 2}^{\rm
i}/M_{\odot}$, $\log (P^{\rm i}/{\rm d})$]=(1.2, 3.5, 0.7) as a
function of the spin-down timescale.} \label{tspm}
  \end{center}
\end{figure}

\subsection{The effect of spin-down timescale on the results}\label{sect:4.1}
In this paper, we assume a spin-down timescale of less than
$10^{\rm 7}$ yr, consistent with the empirical constraints
(\citealt{MENGXC13,MENGXC18}), but the exact value of the
spin-down timescale is very uncertain. Based on the simple
consideration of the conservation of angular momentum,
\citet{MENGXC18b} found that the more massive the initial WD, the
shorter the spin-down timescale. Here, the uncertainty of the
spin-down timescale has a great impact on whether the sdB
companion may be produced and on how common the SNe Ia with the
sdB companions are. Here, we choose a model with the initial
parameters of [$M_{\rm WD}^{\rm i}/M_{\odot}$, $M_{\rm 2}^{\rm
i}/M_{\odot}$, $\log (P^{\rm i}/{\rm d})$]=(1.2, 3.5, 0.7) to
check the effect of the spin-down timescale, where the definition
of the spin-down timescale here is the same to that in section
\ref{sect:2}. Fig.~\ref{hrdspsn} shows the effect of the spin-down
timescale on the evolution of the companion. From the figure, we
can see that if the spin-down timescale is longer than
$3-4\times10^{\rm 6}$ yr, the companion will show the properties
of a sdB star at the moment of supernova explosion.
Fig.~\ref{tspm} shows the final WD and companion masses as a
function of spin-down timescale. We can see from the figure that
the final WD mass does not change with the spin-down timescale if
the spin-down timescale is longer than $10^{\rm 6}$ yr, while the
companion mass does not change with the spin-down timescale until
the spin-down timescale is longer than $3-4\times10^{\rm 6}$ yr,
i.e. although the WD stop increasing its mass, the companion are
still losing its material by mass transfer or stelar wind.
Considering that there is an uncertainty of $\sim10^{\rm 6}$ yr on
the onset of the spin-down phase, the discussion above means that
if the real spin-down timescale is longer than $2-3\times10^{\rm
6}$ yr, our results are not significantly affected by the
spin-down timescale. Via an empirical method, \citet{MENGXC13}
have shown that the spin-down timescale is shorter than a few
$10^{\rm 7}$ yr, and a spin-down timescale of $2-3\times10^{\rm
6}$ yr is consistent with the constraint in \citet{MENGXC13}.

\begin{figure}
\begin{center}
\epsfig{file=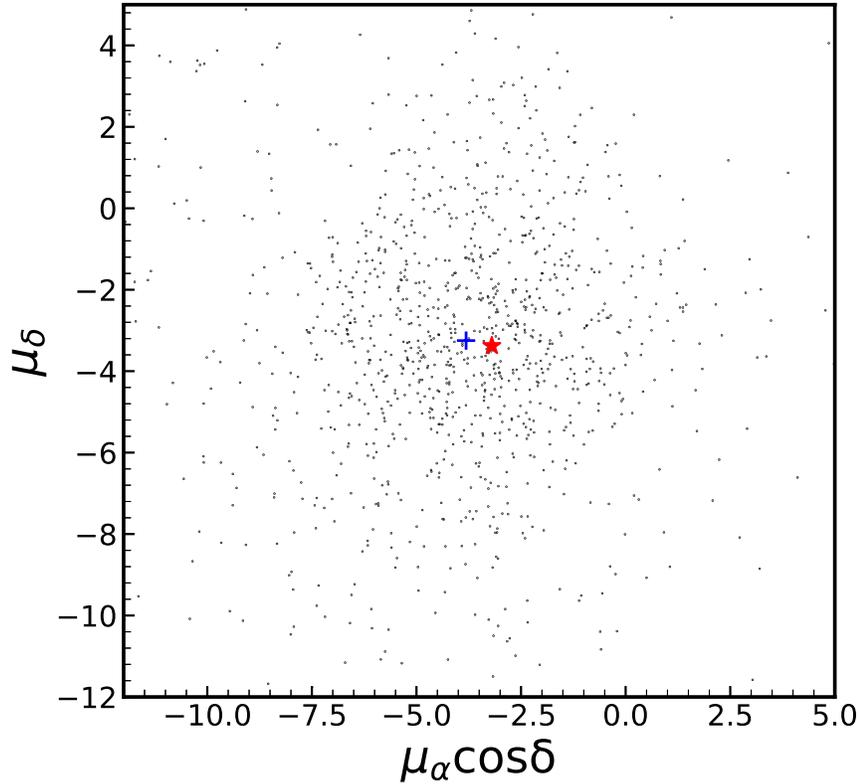,angle=0,width=12.2cm}
 \caption{The proper motions of the stars within $5'$ of the
remnant center of SN 1006. The blue cross presents the median
value of the proper motions of the stars, while the red star
represent the SM stars (\citealt{SCHWEIZER80}).} \label{srnpm}
  \end{center}
\end{figure}

\subsection{The surviving companion in the remnant of SN 1006}\label{sect:4.2}
Searching the surviving companions in SNRs is a powerful way to
distinguish different models. At present, MS, RG and luminous WD
are excluded as the surviving companion of SN 1006
(\citealt{GONZALEZ12}; \citealt{KEERZENDORF12,KEERZENDORF17}), but
sdB stars are not. Interestingly, there is indeed a star similar
to our prediction in the remnant of SN 1006, although it is redder
than our prediction. The star is indeed a single sdB star, i.e.
the Schweizer-Middleditch (SM) star, whose distance is also
consistent with the remnant of SN 1006 (\citealt{SCHWEIZER80};
\citealt{WINKLER03}). Three main effects could contribute to the
redder color of the candidate. 1) Before supernova explosion, the
system experiences a common envelope wind phase. With the
expansion and cooling of the CE material, dust may form around the
progenitor system as observed in some normal and special SNe Ia
(\citealt{LUGL13}; \citealt{FOX13}; \citealt{SILVERMAN13};
\citealt{JOHANSSON13}; \citealt{FOX16}). The dust could redden the
color of the candidate. 2) In this paper, we do not consider the
interaction between supernova ejecta and the companion. The
supernova ejecta may collide into the companion and inject some
kinetic energy of the supernova ejecta into the envelope, and then
huff the envelope (\citealt{MAR00}; \citealt{MENGXC07}). After the
interaction, the envelope will re-establish dynamical equilibrium
soon, but still in a process into thermal equilibrium, lasting for
$10^{\rm 3}-10^{\rm 4}$ yr. During this phase, the effective
temperature of the companion could be lower than that predicted
here by 1000 to 3000 K depending on the ejection energy, i.e. a
slightly redder color (\citealt{PODSIADLOWSKI03};
\citealt{PANK14}). However, the effect could be very small for the
small solid angle of the sdB companions relative to the explosion
center. 3) The dust may also form from the supernova ejecta itself
and such dust may also contributes to a part of color excess
(\citealt{WILLIAMS14}; \citealt{RHO18}). So, a redder color is not
unimaginable for the SM star. Actually, the color excess of the SM
star may be as large as $E(B-V)=0.16\pm0.02$
(\citealt{BURLEIGH00}). If such a color excess is considered, the
color of the SM star is consistent with our prediction. In
addition, the surface gravity and effective temperature of the MS
star are $\log g=6.18\pm0.3$ and $32900\pm340$ K, respectively,
which are also consistent with our results (see Fig.~\ref{grasn7},
\citealt{BURLEIGH00}). Moreover, there is a thin hydrogen envelope
for the SM star as we predicted (\citealt{BURLEIGH00}). Then, if
the SM star is the surviving companion of SN 1006, it must not be
from the WD + He star channel, since no hydrogen on the helium
companion is leaved for the WD + He star channel.

However, due to the presence of red-shifted absorption lines from
SN ejecta, the SM star was suggested to be a background star, and
its distance was suggested to be only slightly greater than the
remnant of SN 1006 (\citealt{WUCC1983};
\citealt{WINKLER03,WINKLER05}). However, the red-shifted
absorption lines could also be derived from an asymmetrical
explosion as indicated by the strongly asymmetric profiles of
Fe$_{\rm II}$ and Si$_{\rm II}$ line and the asymmetric
distribution of the elements in the remnant (\citealt{WINKLER05,
WINKLER11}; \citealt{HAMILTON07}; \citealt{UCHIDA13};
\citealt{ZHOUP18}). The asymmetric explosion could be a common
properties of SNe Ia (\citealt{MAEDA10}).

Therefore, the kinetics characteristics of the star could be the
only piece to judge whether or not the SM star is the surviving
companion of SN 1006. If its space velocity is significantly
different from the other stars in the remnant of SN 1006, the
probability to be the surviving companion would become high.
Otherwise, the probability becomes low. We check the proper motion
of the stars within $5'$ of the remnant center from GAIA DR2, as
shown in Fig.~\ref{srnpm}. From the figure, it seems that there is
not difference between the SM star and other stars in the remnant
of SN 1006 in the aspect of proper motion, i.e. the proper motion
of the SM star only slightly deviates from the median value of the
proper motions of the stars at the direction of the SNR center of
SN 1006, and such a proper motion disfavors the SM star as the
surviving companion of SN 1006 (\citealt{SCHWEIZER80};
\citealt{BURLEIGH00}). So, a 3D space velocity is helpful to judge
the nature of the SM star. However, unfortunately, some data of
the SM star in GAIA DR2 are so uncertain that we cannot use them
to constrain its 3D space velocity, otherwise we could obtain a
complete wrong conclusion\footnote{For example, the effective
temperature of the SM star in GAIA DR2 is definitely not the
effective temperature of a sdB star, i.e. $8869^{\rm +724}_{\rm
-906}$ K, which is derived from a wrong effective temperature
pattern for the stars with the effective temperature of larger
than 10000 K (\citealt{ANDRAE18}).}. For example, the parallax of
the SM star is $\varpi=0.0736\pm0.1244$, and then $\sigma_{\rm
\varpi}/\varpi=1.69$ which is much larger than the threshold value
of $0.2$ for distance estimation from GAIA DR2 data
(\citealt{ASTRAATMADJA16}; \citealt{KATZ18}). The distance of the
SM star from this parallax is much larger than all the previous
measurements from spectrum by at least a factor of $2$ (see
summary in \citealt{BURLEIGH00}). Considering that some other
astrometric measurements of the SM star are also very uncertain,
we applied the measurements in the previous literatures as the
distance of the SM star. Based on a radial velocity of
$-13\pm17~{\rm km^{\rm -1}}$ and a distance of $2.07\pm0.18$ kpc
(\citealt{SCHWEIZER80}; \citealt{WINKLER03};
\citealt{KEERZENDORF17}), we can calculate the $UVW$ velocities of
the SM star, i. e. $U=-5.2\pm14~{\rm km^{\rm -1}}$,
$V=197\pm10~{\rm km^{\rm -1}}$ and $W=3.1\pm5~{\rm km^{\rm -1}}$.
The $V$ value of the SM star is smaller than that of a normal disk
star. We then transform these velocities into the Galactic
rotational velocity at a Galactocentric distance of $\sim6.67$
kpc,
i.e. $V_{\rm c}=196\pm12~{\rm km^{\rm -1}}$, which is smaller than
the Galactic rotational velocity of the disk stars at the
Galactocentric distance by $50\pm19~{\rm km^{\rm -1}}$
(\citealt{HUANGY16}). This velocity difference is marginally
consistent with the predicted orbital velocity here (see
Fig.~\ref{vorr2sn7}). In addition, the smaller rotational velocity
of the SM star may explain its small proper motion shown in
Fig.~\ref{srnpm}. So, the SM star is still possible to be the
surviving companion of SN 1006.

Since we still can not completely exclude the SM star as the
surviving companion of SN 1006, we propose a further detailed
study on the SM star in the remnant of SN 1006. If the candidate
is the surviving companion of SN 1006, it will be a great step for
the progenitor study of SNe Ia. On the contrary, if it is not,
e.g. the future precise distance measurement makes sure that the
SM star is a background star, the survey in the remnant of SN 1006
would favor the DD model or a long spin-down timescale as
suggested in \citet{KEERZENDORF17}, or a hybrid explosion between
core-collapse SN and SNe Ia without the surviving companion
predicted in \citet{MENGXC18}.

\subsection{The surviving companion in the remnant of Kepler's supernova}\label{sect:4.3}
Similarly, no MS or RG star in the remnant of Kepler's supernova
is suitable to be the surviving companion of Kepler's supernova
(\citealt{RUIZLAPUENTE17}). We propose a deeper survey in the
remnant by a blue band to check whether there is a hot subdwarf
star as we predicted. If there is, it will support the CEW model
and then the SD model. Otherwise, the core-degenerate scenario
would be an alternative, i.e. the Kepler's supernova could
originate from the merger of a CO WD and a massive CO core of an
AGB star (\citealt{KASHI11}; \citealt{RUIZLAPUENTE17}), or the
merger of a hybrid CONe WD with a Chandrasekharmass and its MS
companion is another alternative (\citealt{MENGXC18}). Moreover, a
low-mass MS star with a luminosity as dim as 0.2 $L_{\odot}$ can
also not be completely excluded by the observation in
\citet{RUIZLAPUENTE17}.

Whatever, an question should be noticed, i.e. why do the surviving
companions in the both Galactic SNRs tend to be sdB stars? Based
on BPS studies here, as much as 22\% of SNe Ia will explode with
the sdB companions. Actually, to be a sdB star at the moment of
supernova explosion, the mass transfer between the WD and its
companion must begin when the companion is crossing the HG. Some
binary population synthesis studies have shown that most of the
binary systems indeed begin their mass transfer at the HG (e.g.
see Fig. 11 in \citealt{MENG09}), which means that rather a part
of the surviving companions in SNRs could be the sdB stars. The
CEW model even increases the probability that a sdB star is the
surviving companion in a SNR. Compared with the OTW model, the
parameter space leading to SNe Ia from our CEW model extends to
more massive companions and to longer initial orbital period (see
Fig. 7 in \citealt{MENGXC17a}). For the systems in the increased
parameter space, the companions just tend to appear as the sdB
stars at the moment of supernova explosion for a spin-down
timescale assumed here. As shown in Fig.~\ref{hrdsn7}, there is a
connection between the sdB and the RG companions, i.e. if the
spin-down timescale is longer than the timescale assumed here,
more companion will show the properties of the sdB stars. This
could explain the negative results for searching the surviving
companion in most of SNRs, since the surveys usually focus on
luminous MS or RG candidate.

Since a sdB star has a life much longer than a SNR, we would
expect that there are many sdB stars from SNe Ia wandering in the
Galaxy. In the future, we will give a detailed study on such sdB
stars from our SN Ia progenitor model.\\
\\
In summary, based on the CEW model in \citet{MENGXC17a}, we study
the properties of the companions at the moment of supernova
explosion in details, where we assume a spin-down timescale of
less than $10^{\rm 7}$ yr. We found that the companions may be the
MS, RG and sdB stars, although the initial systems are WD + MS
ones. Our results may explain the observations constraining the
progenitors of some SNe Ia. Here, we suggest for the first time
that the sdB stars as the surviving companions of SNe Ia may be
from the WD + MS system. Especially, the sdB companions here are
quite different from the helium companions from the WD + He star
channel. The SNe Ia with the sdB companions may contribute as much
as 22\% to all SNe Ia. We propose a deep search for the surviving
companions in some supernova remnants in $U$ or $UV$ bands.

\section*{Acknowledgments}

This work was partly supported by the NSFC (No. 11473063,
11522327, 11390374, 11521303 and 11733008), Yunnan Foundation (No.
2015HB096, 2017HC018), CAS light of West China Program and CAS
(No. KJZD-EW-M06-01).


\label{lastpage}
\end{document}